\documentclass[aps,pre,superscriptaddress,showpacs,floatfix]{revtex4-1}
\usepackage{amsfonts}
\usepackage{graphicx,amsmath,amssymb,xspace,epsfig,float,multirow,subfigure,tabularx}
\usepackage{hyperref}

\begin{document}

\title{Covariant Bethe - Salpeter approximation in strongly correlated
electron systems model}
\author{Zhenhao Fan}
\affiliation{School of Physics, Peking University, Beijing 100871, China}
\affiliation{Collaborative Innovation Center of Quantum Matter, Beijing,
China}
\author{Zhipeng Sun}
\affiliation{School of Physics, Peking University, Beijing 100871, China}
\affiliation{Collaborative Innovation Center of Quantum Matter, Beijing,
China}
\author{Dingping Li}
\email{lidp@pku.edu.cn}
\affiliation{School of Physics, Peking University, Beijing 100871, China}
\affiliation{Collaborative Innovation Center of Quantum Matter, Beijing,
China}

\author{Itzhak Berenstein}
\email{beatseek@live.com}
\affiliation{Department of Computer Science£¬Open University£¬Raanana 43107£¬ Israel}
\author{Guy Leshem}
\email{gleshem2525@gmail.com}
\affiliation{Computer Center,Ben Gurion University, Be'er Sheva, 84105, Israel}

\author{Baruch Rosenstein}
\email{baruchro@hotmail.com}
\affiliation{Electrophysics Department, National Chiao Tung University, Hsinchu 30050,
\textit{Taiwan, R. O. C}}

\begin{abstract}
Strongly correlated electron systems are generally described by tight
binding lattice Hamiltonians with strong local (on site) interactions, the
most popular being the Hubbard model. Although the half filled Hubbard model
can be simulated by Monte Carlo(MC), physically more interesting cases beyond
half filling are plagued by the sign problem. One therefore should resort to
other methods. It was demonstrated recently that a systematic truncation of
the set of Dyson - Schwinger equations for correlators of the Hubbard,
supplemented by a \textquotedblleft covariant" calculation of correlators
leads to a convergent series of approximants. The covariance preserves all
the Ward identities among correlators describing various condensed matter
probes. While first order (classical), second (Hartree - Fock or gaussian)
and third (Cubic) covariant approximation were worked out, the fourth
(quartic) seems too complicated to be effectively calculable in fermionic
systems. It turns out that the complexity of the quartic calculation\ in
local interaction models,is manageable computationally. The quartic (Bethe -
Salpeter type) approximation is especially important in 1D and 2D models in
which the symmetry broken state does not exists (the Mermin - Wagner
theorem), although strong  fluctuations dominate the physics at strong
coupling. Unlike the lower order approximations, it respects the Mermin -
Wagner theorem. The scheme is tested and exemplified on the single band 1D
and 2D Hubbard model.
\end{abstract}

\maketitle

\section{Introduction}

Strongly correlated electron systems like the high critical temperature
cuprate superconductors\cite{dagotto1994correlated,scalapino2012common},
quantum magnets\cite{Fradkin}, etc. are currently a topic of great interest
in condensed matter physics. These materials, both three dimensional (3D),
layered or recently fabricated 2D materials\cite{Novoselov} typically (but
not always) involve hybridization of the $d$ or $f$ \ atomic orbitals.
Unfortunately understanding of physics of this class of materials hinges on
at least qualitative understanding of the simplest models like the one band
Hubbard model on square lattice. Although the Hamiltonian of the model,
often just a nearest neighbor tunneling and strong (large $U$) local
(on site) interactions, is deceptively simple, exact solution\cite{Lieb-Wu}
exists only in 1D. Moreover even in 1D it is limited to specific quantities like the
distribution of momenta. Green's functions that are directly
measured in ARPES experiments or susceptibilities measured in magnetization
or optical experiments have not been calculated exactly. Alternatively one
can solve yet smaller Hubbard -like systems on periodic \textquotedblleft
crystallites" around $16$ sites by exact diagonalization\cite{ED}. Therefore
one has to resort to approximations of various kinds.

A straightforward  approach is the path integral Monte Carlo\cite{MC}
simulation of fermionic systems. Unfortunately the sign problem
immediately arises in the case of interests. The simple half filled Hubbard
model has no sign problem due to the electron hole symmetry; however, for
example the high temperature superconductivity occurs at nonzero doping,
when the electron - hole symmetric condition is violated. In most cases of
interest the sign problem therefore does not allow simulation. An
alternative is a diagrammatic approximation scheme or some other analytic
methods. Most start with an (infinite) hierarchy of relations between
correlators known as Dyson - Schwinger equations (DSE). This infinite system
of generally nonlinear equations should be somehow disentangled. This can be
done either by devising some kind of perturbation theory (weak, strong
coupling, large \textquotedblleft $N$" expansion etc) or directly by truncating the
equations according to some \textquotedblleft principle", like variational or
other\cite{albrecht1998ab, onida2002electronic}.

The quality of a truncation is usually dependent on preserving general
relations like conservation laws and \textquotedblleft sum rules"\cite%
{baym1961conservation, haussmann2007thermodynamics}. This is highly
nontrivial and several strategies were attempted\cite{benchmarkref}. A
general method to preserve the Ward - Takahashi identity(WTI) in an
approximation scheme was developed long time ago\cite{Kovner} in the context
of field theory as the covariant gaussian approximation (CGA) to solve
unrelated problems in quantum field theory and superfluidity\cite%
{Kovner2,Kovner3,LiQiong,ZhangYahui}. A non-perturbative variational
gaussian method that originated in quantum mechanics of atoms and molecules
in relativistic theories like the standard model of particle physics has
several serious related problems. First, the wave function renormalization
requires a dynamical description. Second, the Green's functions obtained
using the naive gaussian approximation violates the charge conservation. In
particular the most evident problem is that the Goldstone bosons resulted
from spontaneous breaking of continuous symmetry are massive. The method is
thus considered dubious or inconsistent. Both problems were solved by an
observation that the solutions of the minimization equations are not
necessarily equivalent to the variational Green's functions. This results in
the covariant gaussian approximation (CGA).

It was demonstrated recently\cite{Cubic} (using a simple example) that a
systematic truncation of the set of Dyson - Schwinger equations for
correlators of the Hubbard, supplemented by a \textquotedblleft covariant"
calculation of correlators, leads to a converging series of approximants.
The covariance preserves all the Ward identities among correlators
describing various condensed matter probes. While first order (classical)
second order (Hartree - Fock or gaussian) and third order (cubic) approximations have been
worked out, the fourth order seems too complicated to be effectively calculable
in fermionic systems without symmetry breaking. This is especially important
in 1D and 2D models in which the symmetry broken state does not exist (due
to the Mermin - Wagner theorem\cite{mermin1966absence,mermin1967absence}),
although strong anti - ferromagnetic (AF) fluctuations dominate the physics
at strong coupling. Unlike lower order approximations, the quartic covariant
scheme already respects the Mermin - Wagner theorem.

It is shown in the present paper that the complexity of \emph{covariant
quartic approximation }(CQA) in Hubbard model can be reduced to a manageable
level due to locality of interactions. There is a possibility of transitions
between the coordinate and the momentum spaces that reduces the computation
cost. We focus on the electron correlator describing the electron (hole)
excitations measured in photoemission and other condensed matter probes. The
scheme is tested and exemplified on solvable 1D finite site Hubbard model
(including beyond half filling) and on a more physically important 2D
Hubbard model at half filling in which reliable Monte Carlo simulations
exist.

The paper is organized as follows. In section \ref{sec:hierarchy} the
sequence of covariant approximations is developed using the simplest
possible quantum model: the one dimensional quantum anharmonic oscillator.
Next in section \ref{sec:generalformula} the CQA is developed for a general
fermionic system and is applied to the one band Hubbard model in Section \ref%
{sec:Hubbardformula}. In Section \ref{sec:implement}, the
implementation is described and the results are compared with
exact solutions(1D) and Monte Carlo results(2D).
An estimate of complexity of application of CQA to a realistic 2D material are
subject of Section \ref{sec:complexity}. Discussion and conclusions are presented
in Section \ref{sec:discussion}.

\section{Hierarchy of covariant truncations of Dyson - Schwinger equations
in a bosonic model}

\label{sec:hierarchy}

The main idea and methodology behind the covariant approximants are
presented in this section in the simplest possible setting: thermodynamics
of the bosonic field (often used as an ``order parameter"\cite%
{amit2005field, chaikin1995principles}). Later the most advanced fourth in a
series of such approximant for a many - body fermionic system will be
developed and shown to be computationally practical.

\subsection{An exactly solvable ``bosonic" model: 1D classical statistical
system (quantum mechanics of the 1D anharmonic oscillator)}

The simplest nontrivial model having many of the basic ingredients of the
interacting electronic system is the one dimensional scalar field theory
viewed as the thermodynamic 1D Ginzburg - Landau -Wilson model\cite%
{amit2005field, chaikin1995principles} (equivalently via Feynman path
integral\cite{NO} to the quantum mechanics of the anharmonic oscillator).
The quantity describing the thermal fluctuations of the field is the
Boltzmann factor:

\begin{equation}
F\left[ J\right] \equiv H\left[ J\right] /T=\int_{x=-\infty }^{\infty }\left
\{ -\frac{1}{2}\psi _{x}\partial ^{2}\psi _{x}+\frac{a}{2}\psi _{x}^{2}+%
\frac{1}{4}\psi _{x}^{4}-J_{x}\psi _{x}\right \} .  \label{Hamiltonian}
\end{equation}%
The scale and the field normalization are chosen in a way to make the
coefficient of the gradient term to be $1/2$ and that of the interaction
term $1/4$. Here $J_{x}$ is an external ``source", a convenient theoretical
device to generate connected correlators\cite%
{amit2005field,chaikin1995principles}
\begin{eqnarray}
W\left[ J\right] &=&-\ln \int_{\psi }e^{-F\left[ J\right] }  \label{Gn} \\
G_{x_{1}...x_{n}} &=&-\frac{\delta }{\delta J_{x_{1}}}...\frac{\delta }{%
\delta J_{x_{n}}}W\left[ J\right] \equiv \left \langle \psi _{x_{1}}...\psi
_{x_{n}}\right \rangle  \notag
\end{eqnarray}

\subsection{The main idea of the approximation}

The CQA is derived following the line of reasoning used to obtain other
covariant approximations, gaussian and cubic in Refs. %
\onlinecite{Wang17,Cubic}. It is based on the set of Dyson-Schwinger
equations(DSE). Here however we go one step further by considering the
truncation of DSE up to the four field correlator. The hierarchy of DSE for
the connected correlators contains, the first equation for a nonvanishing
source,

\begin{equation}
J_{x}=\left( -\partial ^{2}+a\right) \varphi _{x}+\varphi _{x}^{3}+3\varphi
_{x}G_{xx}+G_{xxx};  \label{eqm}
\end{equation}%
also called the ``equation of motion". Here the field expectation value in
the presence of the source is denoted by $\varphi _{x}\equiv \left \langle
\psi _{x}\right \rangle $. The second equation, obtained by the functional
differentiation with respect to the source, is:
\begin{align}
\delta \left( x_{1}-x_{2}\right) & =\left( -\partial _{x_{1}}^{2}+a+3\varphi
_{x_{1}}^{2}\right) G_{x_{2}x_{1}}+3G_{x_{2}x_{1}}G_{x_{1}x_{1}}
\label{gapeqcon} \\
& +3\varphi _{x_{1}}G_{x_{2}x_{1}x_{1}}+G_{x_{2}x_{1}x_{1}x_{1}}\text{.}
\notag
\end{align}%
The third equation,
\begin{align}
0& =\left( -\partial _{x_{1}}^{2}+a+3\varphi
_{x_{1}}^{2}+3G_{x_{1}x_{1}}\right) G_{x_{3}x_{2}x_{1}}+6\varphi
_{x_{1}}G_{x_{3}x_{1}}G_{x_{2}x_{1}}  \label{cubic} \\
&
+3G_{x_{2}x_{1}}G_{x_{3}x_{1},x_{1}}+3G_{x_{3}x_{1}}G_{x_{2}x_{1}x_{1}}+3%
\varphi _{x_{1}}G_{x_{3}x_{2}x_{1},x_{1}}+G_{x_{3},x_{2},x_{1},x_{1},x_{1}},
\notag
\end{align}%
if written in full. The last term is the five point correlator and thus will
be omitted within the fourth order approximation,
\begin{align}
0& =\left( -\partial _{x_{1}}^{2}+a+3\varphi
_{x_{1}}^{2}+3G_{x_{1}x_{1}}\right) G_{x_{3}x_{2}x_{1}}+6\varphi
_{x_{1}}G_{x_{3}x_{1}}G_{x_{2}x_{1}}  \label{cubic1} \\
&
+3G_{x_{2}x_{1}}G_{x_{3}x_{1}x_{1}}+3G_{x_{3}x_{1}}G_{x_{2}x_{1}x_{1}}+3%
\varphi _{x_{1}}G_{x_{3}x_{2}x_{1}x_{1}},  \notag
\end{align}%
The fourth DSE is already too cumbersome to write in full. Here only even
correlators are retained (with correlators up to the four field):
\begin{eqnarray}
0 &=&\left( -\partial _{x_{1}}^{2}+a+3G_{x_{1}x_{1}}\right)
G_{x_{4}x_{3}x_{2}x_{1}}+6G_{x_{4}x_{1}}G_{x_{3}x_{1}}G_{x_{2}x_{1}}+3G_{x_{2}x_{1}}G_{x_{4}x_{3}x_{1}x_{1}}
\label{quarticeq} \\
&&+3G_{x_{3}x_{1}}G_{x_{4}x_{2}x_{1}x_{1}}+3G_{x_{4}x_{1}}G_{x_{3}x_{2}x_{1}x_{1}}+...%
\text{.}  \notag
\end{eqnarray}%
The ``..." refers to terms containing odd correlators, namely $\varphi _{x}$
or three field correlators. In the next subsection it will be argued that
symmetry makes odd correlators redundant.

The set of equations continues to higher orders and will become too
complicated. However at least far from second order phase transition point
(criticality) it is natural to assume a clustering hypothesis: higher
connected correlators are \textquotedblleft smaller" in most models
describing (even strongly coupled) physical systems. The covariant
truncation is one of the proposals to truncate the infinite set of
increasingly complicated equations using the symmetry and consistency
arguments.

\subsection{The truncation of the DSE set and the symmetry considerations}

The ``quartic covariant" truncation of the infinite set of equations is
achieved by taking $G_{x_{1}...x_{i}}=0,$ for all $i>4$. The set for $%
J_{x}=0 $

\begin{gather}
0=\left( -\partial ^{2}+a\right) \varphi _{x}+\varphi _{x}^{3}+3\varphi
_{x}G_{xx}^{tr}+G_{xxx}^{tr};  \label{minimeqs} \\
\delta \left( x_{1}-x_{2}\right) =\left( -\partial _{x_{1}}^{2}+a+3\varphi
_{x_{1}}^{2}\right)
G_{x_{2}x_{1}}^{tr}+3G_{x_{2}x_{1}}^{tr}G_{x_{1}x_{1}}^{tr}+3\varphi
_{x_{1}}G_{x_{2}x_{1}x_{1}}^{tr}+G_{x_{2}x_{1}x_{1}x_{1}}^{tr};  \notag \\
0=\left( -\partial _{x_{1}}^{2}+a+3\varphi _{x_{1}}^{2}\right)
G_{x_{3}x_{2}x_{1}}^{tr}+6\varphi
_{x_{1}}G_{x_{3}x_{1}}^{tr}G_{x_{2}x_{1}}^{tr}+3G_{x_{1}x_{1}}^{tr}G_{x_{3}x_{2}x_{1}}^{tr}+3G_{x_{2}x_{1}}^{tr}G_{x_{3}x_{1}x_{1}}^{tr}+3G_{x_{3}x_{1}}^{tr}G_{x_{2}x_{1}x_{1}}^{tr}+3\varphi _{x_{1}}G_{x_{3}x_{2}x_{1}x_{1}}^{tr};
\notag \\
0=\left( -\partial _{x_{1}}^{2}+a\right)
G_{x_{4}x_{3}x_{2}x_{1}}^{tr}+6G_{x_{4}x_{1}}^{tr}G_{x_{3}x_{1}}^{tr}G_{x_{2}x_{1}}^{tr}+3G_{x_{1}x_{1}}^{tr}G_{x_{4}x_{3}x_{2}x_{1}}^{tr}+3G_{x_{2}x_{1}}^{tr}G_{x_{4}x_{3}x_{1}x_{1}}^{tr}+3G_{x_{3}x_{1}}^{tr}G_{x_{4}x_{2}x_{1}x_{1}}^{tr}+3G_{x_{4}x_{1}}^{tr}G_{x_{3}x_{2}x_{1}x_{1}}^{tr}+...%
\text{,}  \notag
\end{gather}%
will be called the ``minimization equations". They will determine the
``truncated" correlators that should be distinguished\cite%
{Kovner,Wang17} from \ an approximant to the ``physical" correlators defined
in Eq.(\ref{Gn}).

Moreover in many cases symmetry simplifies the set of equations. In our case
the Hamiltonian, Eq.(\ref{Hamiltonian}) has the global $Z_{2}$ symmetry $%
\psi_{x}\rightarrow-\psi_{x}$ that will be preserved in the covariant
approach. The symmetry allows to conclude that $%
\varphi_{x}^{tr}=G_{x_{1},x_{2},x_{3}}^{tr}=0$. Then the first and third
equations are automatically satisfied, and the remaining second and fourth
will be simplified to

\begin{equation}
\delta \left( x_{1}-x_{2}\right) =\left( -\partial _{x_{1}}^{2}+a\right)
G_{x_{2}x_{1}}^{tr}+3G_{x_{2}x_{1}}^{tr}G_{x_{1}x_{1}}^{tr}+G_{x_{2}x_{1}x_{1}x_{1}}^{tr};
\label{GE}
\end{equation}

\begin{eqnarray}
&&\left( -\partial _{x_{1}}^{2}+a\right)
G_{x_{4}x_{3}x_{2}x_{1}}^{tr}+6G_{x_{4}x_{1}}^{tr}G_{x_{3}x_{1}}^{tr}G_{x_{2}x_{1}}^{tr}+3G_{x_{1}x_{1}}^{tr}G_{x_{4}x_{3}x_{2}x_{1}}^{tr}
\label{BS} \\
&&+3G_{x_{2}x_{1}}^{tr}G_{x_{4}x_{3}x_{1}x_{1}}^{tr}+3G_{x_{3}x_{1}}^{tr}G_{x_{4}x_{2}x_{1}x_{1}}^{tr}+3G_{x_{4}x_{1}}^{tr}G_{x_{3}x_{2}x_{1}x_{1}}^{tr}=0%
\text{.}  \notag
\end{eqnarray}

The first equation is commonly called ``gap equation" due to similarity with
the corresponding equation in the superconductivity theory, while the second
is similar, but not equivalent to the ``Bethe - Salpeter" equation for the
bound states. We address this issue later after using the translation
invariance of the equations.

\subsection{Covariant vs ``naive" correlator}

The connected correlators are obtained by differentiations of the field
shift $\varphi _{x}$ with respect to the sources, and subsequently taken at $%
J=0$,
\begin{equation}
G_{x_{1}...x_{n}}\equiv \frac{\delta }{\delta J_{x_{1}}}...\frac{\delta }{%
\delta J_{x_{n-1}}}\varphi _{x_{n}}|_{J_{x_{i}}=0}\text{,}  \label{Gdef}
\end{equation}%
where $\varphi _{x_{n}}$ is given by the solution via Eqs.(\ref{eqm},\ref%
{gapeqcon},\ref{cubic1},\ref{quarticeq}). The correlators obtained in that
way are not necessarily equal to truncated functions that appear in the
minimization equations Eq.(\ref{minimeqs}). Sometimes $G^{tr}$ in the
minimization equations are treated as approximate correlators both for $I=2$
(gaussian) and $I=4$ (Bethe - Salpeter). We will refer to these as
\textquotedblleft naive" noncovariant approach. In these cases one often
discovers that a symmetry, like the $Z_{2}$ are not respected. Generally it
is found that so called Ward identities (relations between various Green's
functions derived from the symmetry) are not obeyed by the approximate
correlators. In extreme cases, for example when a continuous symmetry is
spontaneously \ broken, Ward identities like the Goldstone theorem are
violated. As will be discussed below, even the antisymmetry of the fermionic
correlators (the Fermi symmetry) is violated in the naive truncation
approaches. One expects that the covariant approximation results are more
accurate.

The method was compared with available exact results for the S-matrix in the
Gross - Neveu model\cite{GN} (a local four Fermion interactions in 1D Dirac
excitations recently considered in condensed matter physics) and with MC
simulations in various scalar models, see Ref.\onlinecite{Wang17} . Applied
to the electronic field correlator in electronic systems, CGA becomes
roughly equivalent to Hartree - Fock (HF) approximation that is generally
not precise enough. Its covariance might improve the calculation of the four
fermion correlators like the density - density, but to address
quantitatively photoemission or other direct electron or hole excitation
probes, a more precise method is needed.

The covariant correlators $G_{x_{1}...x_{n}}$ could be different from $%
G_{x_{1}...x_{n}}^{tr}$, and the covariant correlators satisfy the Ward
identities \cite{Wang17,Cubic}. In our case, differentiating Eq.(\ref{eqm}),
one obtains:
\begin{equation}
\delta \left( x_{1}-x_{2}\right) =\left( -\partial _{x_{1}}^{2}+a+3\varphi
_{x_{1}}^{2}+3G_{x_{1}x_{1}}^{tr}\right) G_{x_{2}x_{1}}+3\varphi _{x_{1}}%
\frac{\delta G_{x_{1}x_{1}}^{tr}}{\delta J_{x_{2}}}+\frac{\delta
G_{x_{1}x_{1}x_{1}}^{tr}}{\delta J_{x_{2}}}\text{.}  \label{fulcorr}
\end{equation}%
Of course in 1D symmetry is not spontaneously broken and thus $\varphi _{x}$
vanishes, so that the covariant correlator obeys
\begin{equation}
\delta \left( x_{1}-x_{2}\right) =\left( -\partial
_{x_{1}}^{2}+a+3G_{x_{1}x_{1}}^{tr}\right) G_{x_{2}x_{1}}+\frac{\delta
G_{x_{1}x_{1}x_{1}}^{tr}}{\delta J_{x_{2}}}  \label{fullcorr1}
\end{equation}%
The derivative $\frac{\delta G_{x_{1}x_{2}x_{3}}^{tr}}{\delta J_{x_{4}}}$,
often called the ``chain correction" (due to its diagrammatic interpretation%
\cite{Kovner}), is obtained by differentiating the third DS equation, Eq.(%
\ref{cubic1}) with respect to $J$ (and omitting the odd correlators, as
above):
\begin{eqnarray}
&&\left( -\partial _{x_{1}}^{2}+a+3G_{x_{1}x_{1}}^{tr}\right) \frac{\delta
G_{x_{3}x_{2}x_{1}}^{tr}}{\delta J_{x_{4}}}+3G_{x_{2}x_{1}}^{tr}\frac{\delta
G_{x_{3}x_{1}x_{1}}^{tr}}{\delta J_{x_{4}}}+3G_{x_{3}x_{1}}^{tr}\frac{\delta
G_{x_{2}x_{1}x_{1}}^{tr}}{\delta J_{x_{4}}}  \label{chaineq} \\
&=&-6G_{x_{4}x_{1}}G_{x_{3}x_{1}}^{tr}G_{x_{2}x_{1}}^{tr}-3G_{x_{4}x_{1}}G_{x_{3}x_{2}x_{1}x_{1}}^{tr}%
\text{.}  \notag
\end{eqnarray}%
Note that the chain equation is generally linear in chain variable. This is
crucial for an ability to calculate the approximation. It will be
demonstrated in the next section that in the particular case of the local
scalar model (in unbroken symmetry phases only), the truncated and the full
two field correlators (or so called the green functions) in fact coincide.
This general observation simplifies the calculation of the correlator since
one just needs to solve the minimization equations.

\subsection{Translation invariance and solution of minimization equations}

We consider in the present paper only translation symmetric phases. The
translation symmetry greatly simplifies the solution of the nonlinear
minimization equations by making the Fourier transformation of the Green
functions. In our case only two are required:
\begin{eqnarray}
G_{x_{1}x_{2}}^{tr} &=&\frac{1}{2\pi }\int_{k}g_{k}^{tr}e^{-ik\left(
x_{1}-x_{2}\right) },  \label{Fourier} \\
G_{x_{4}x_{3}x_{2}x_{1}}^{tr} &=&\frac{1}{\left( 2\pi \right) ^{3}}%
\int_{k_{1}k_{2}k_{3}}g_{k_{1}k_{2}k_{3}}^{tr}e^{-ik_{1}\left(
x_{1}-x_{2}\right) }e^{-ik_{2}\left( x_{2}-x_{4}\right) }e^{-ik_{3}\left(
x_{3}-x_{4}\right) }\text{.}  \notag
\end{eqnarray}%
We denote $g_{k}^{tr}$ as $g_{k}$ because the covariant green function is
equal to the two field correlator solution of the minimization equations.
The equations Eq.(\ref{GE}) and Eq.(\ref{BS}) then take a form:
\begin{eqnarray}
\left( k^{2}+a+3b\right) g_{k}+\frac{1}{2\pi }\text{$\int $}%
_{k_{1}}q_{kk_{1}} &=&1;  \label{kspace} \\
\left( \left( k_{1}+k_{2}+k_{3}\right) ^{2}+a+3b\right)
g_{k_{1}k_{2}k_{3}}^{tr}+3g_{k_{3}}q_{k_{1}k_{2}}+3g_{k_{2}}q_{k_{1}k_{3}}+3g_{k_{1}}q_{k_{2}k_{3}} &=&-6g_{k_{1}}g_{k_{2}}g_{k_{3}}%
\text{,}  \notag
\end{eqnarray}%
where the ``bubble" $b$ and the ``vertex" $q$ are defined as
\begin{equation}
b\equiv G_{xx}=\frac{1}{2\pi }\int_{k}g_{k};\text{ \  \ }  \label{defb}
\end{equation}%
\begin{equation}
q_{k_{1}k_{2}}\equiv \frac{1}{2\pi }\int_{k}g_{k_{1}k_{2}k}^{tr}\text{.}
\label{defq}
\end{equation}

This nonlinear set of equations can be simplified as follows. Dividing the
second Eq.(\ref{kspace}) by $\left(\left(k_{1}+k_{2}+k_{3}\right)^{2}+a+3b%
\right)$, and integrating over $k_{3}$, one obtains a linear equation in $q$:

\begin{equation}
q_{k_{1}k_{2}}=-\frac{3}{2\pi }\int_{k}\left \{ \frac{%
g_{k}q_{k_{1}k_{2}}+g_{k_{2}}q_{k_{1}k}+g_{k_{1}}q_{k_{2}k}}{\left(
k_{1}+k_{2}+k\right) ^{2}+a+3b}+\frac{2g_{k_{1}}g_{k_{2}}g_{k}}{\left(
k_{1}+k_{2}+k\right) ^{2}+a+3b}\right \} \text{.}  \label{vertexsol}
\end{equation}%
The first equation of Eq.(\ref{kspace}) in the form,
\begin{equation}
g_{k_{1}}=\left( k^{2}+a+3b\right) ^{-1}\left( 1-\frac{1}{2\pi }%
\int_{k}q_{k_{1}k}\right) \text{,}  \label{gtr_iter}
\end{equation}%
Eq.(\ref{defb}),Eq.(\ref{gtr_iter}) and Eq.(\ref{vertexsol}) are finally
solved by iterations. The results are presented next.

\subsection{Comparison with exact results.}

\subsubsection{Numerical solution of the minimization equation}

In order to solve the minimization equations, we chose a cut-off $k_{\Lambda
}$,and assume that $g\left( k\right) =1/k^{2}$ for $\left \vert
k\right
\vert >k_{\Lambda }$ and $q\left( k_{1},k_{2}\right) =0$ for $%
\left
\vert k_{1}\right \vert >k_{\Lambda }$ or $\left \vert
k_{2}\right
\vert >k_{\Lambda } $, and the assumption is justified by the
asymptotical forms of $g\left( k\right) $ and $q\left( k_{1},k_{2}\right) $.
Thus
\begin{equation}
b=\frac{1}{2\pi }\int_{-\infty }^{\infty }g\left( k\right) dk=\frac{1}{2\pi }%
\int_{-k_{\Lambda }}^{k_{\Lambda }}g\left( k\right) dk+\frac{1}{\pi
k_{\Lambda }},
\end{equation}%
and other integrals can be approximated by replacing the integral bounds $%
\int_{-\infty }^{\infty }$ by the integration inside the cut-off $k_{\Lambda
}$, $\int_{-k_{\Lambda }}^{k_{\Lambda }}$ plus $\int_{k_{\Lambda }}^{\infty
}+\int_{-\infty }^{-k_{\Lambda }}$ using the asymptotical forms of $g\left(
k\right) $ and $q\left( k_{1},k_{2}\right) $ .We enlarge $k_{\Lambda }$,
until the final result converges. As for the finite integral $%
\int_{-k_{\Lambda }}^{k_{\Lambda }}$, we first split the integration
interval into $N$ intervals and evaluate the value of each intervals using
Bode's rule\cite{NR}, then we split the interval into $2N$ intervals and do
it again until it converges. For the parameters we calculate, they all
converge at cut-off $k_{\Lambda }=102.4$ and intervals $N=1024$.

\subsubsection{Comparison with the exact result}

\begin{figure}[tbp]
\centering
\includegraphics[width=\linewidth]{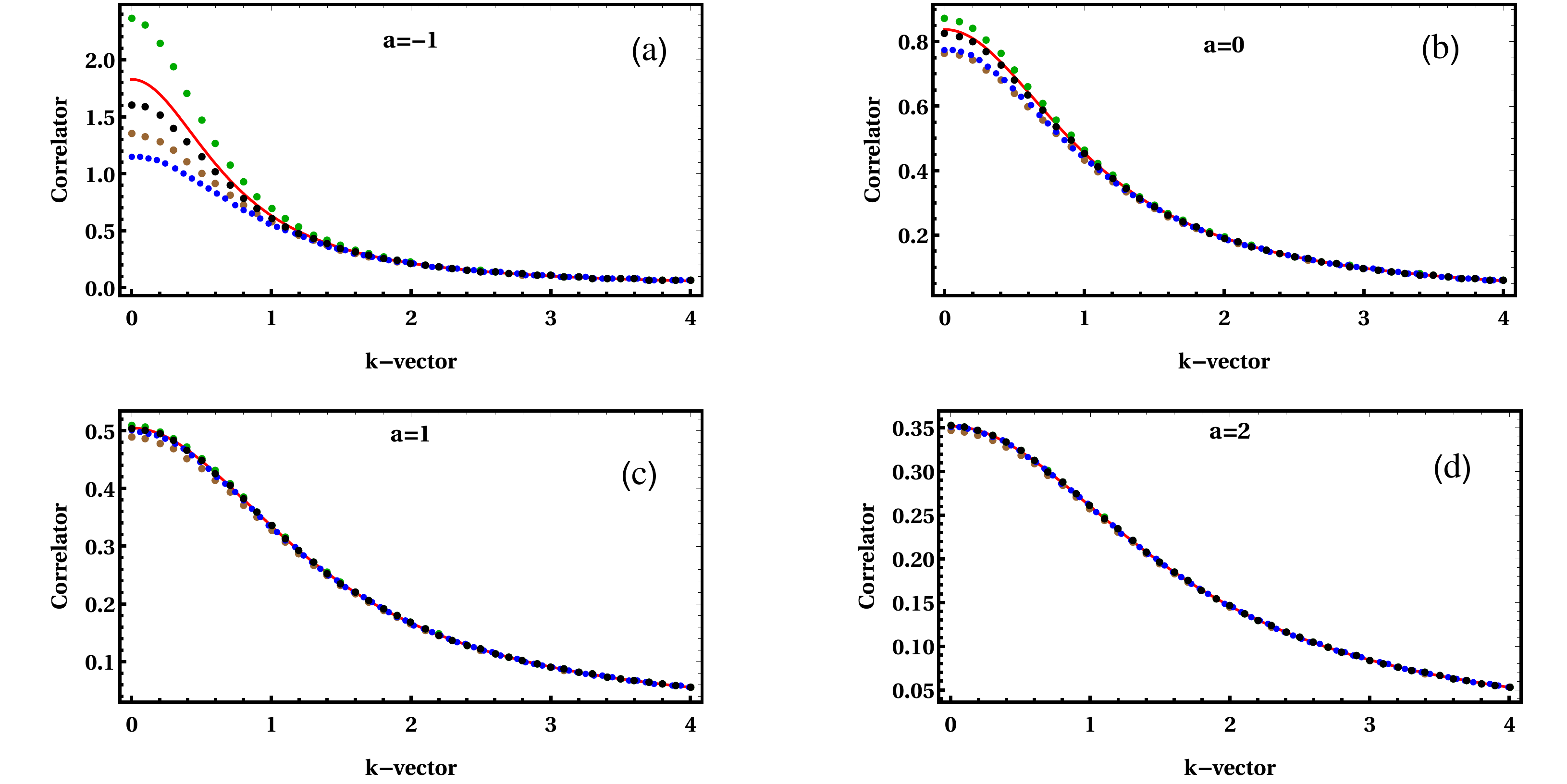}
\caption{Comparison of GW, covariant gaussian approximation (CGA), covariant
cubic approximation (CCA),covariant quartic approximation (CQA), and exact
results for a 1D Ginzburg - Landau - Wilson chain (quantum mechanical
anharmonic oscillator) $a=$ $-1,0.5,0,1$ respectively shown in
(a),(b),(c),(d). The red line is the exact correlator, while blue, brown,
the darker green, and black dots, are GW, CGA, CCA and CQA respectively.}
\label{fig:scalar}
\end{figure}

The exact result can be obtained by analogue with the quantum anharmonic
oscillator \cite{Wang17,Cubic}. We calculate $g_{k}$ for $a=$ $-1,0.5,0,1$
in Fig. \ref{fig:scalar} (the exact is in the red line ) and compare them
with the results obtained by using GW, covariant gaussian approximation
(CGA), covariant cubic approximation (CCA) and covariant quartic approximation
(CQA). For example the worst precision of CQA at $a=0$ is around $1\%$ for
the whole range of $k$ - vectors. For negative $a$, for example $a=-1$, the
deviation is around $12\%$ at $k=0$ for CQA.

\section{Quartic approximation in strongly interacting electronic systems}

\label{sec:generalformula}

Generalization to many - body electron system within the field theoretical
path integral framework is rather straightforward: the field becomes a
Grassmannian function of (Matsubara) time, space and some other ``flavours"
like spin and valley indices. Still the Pauli principle makes the
calculation simpler as is emphasized below in full generality. For example
there is no expectation values of products of odd number of fermionic fields.

\subsection{Four fermion interactions in the Nambu representation and DS
equations.}

Let us start with a general model of fermions described within the path
integral approach\cite{NO} by a large number of complex Grassmannian
variables $\psi ^{a}$ and $\psi ^{\ast a}$. All the usual indices like
position in space, time, spin, band (valley), etc are lumped together in one
``index" $a$. It is convenient to consider the two
conjugate Grassmannians as real Grassmann numbers with an additional\ index $%
c $ taking two values $c=.$ (no star) and$\ c=\ast $. Then all the indices
contained in $a$ and the charge (Nambu) index appear in the combined index $%
A=\left \{ c,a\right \} $ on the same footing. The interaction is assumed to
be of the two - body (four Fermi) variety common to the many - body
electronic systems. The Matsubara action including the Grassmannian source
generally has a form:
\begin{equation}
A\left[ \psi ,J\right] =\frac{1}{2}\psi ^{A}T^{AB}\psi ^{B}+\frac{1}{4!}%
V^{ABCD}\psi ^{A}\psi ^{B}\psi ^{C}\psi ^{D}-J^{A}\psi ^{B}\text{.}
\label{Matsubara_action}
\end{equation}%
The antisymmetric matrix $T$ will be referred to as the ``hopping" term. The
coefficient the of interaction term, $V^{ABCD}$, is also antisymmetric in
all four indices due to the Pauli symmetry.

General connected correlators are defined (similar to scalar
fields, but note that the order of derivatives matters due to the Grassmannian algebra) by,

\begin{equation}
G^{A_{1}A_{2}...A_{n}}=-\frac{\delta ^{n}F}{\delta J^{A_{1}}...\delta
J^{A_{n}}}=\left \langle \psi ^{A_{1}}...\psi ^{A_{n}}\right \rangle \text{,}
\label{Cumulants}
\end{equation}%
where $F=-\log Z$ is the free energy with $Z=\int e^{-A\left[ \psi ,J%
\right] }$.


As with the scalar fields we start from derivation of the relevant DS
equations via differentiation with respect to the source $J$. Then the
covariant correlator is shown to be equal to the ``truncated" one within a
specific form of the minimization equations. The first in the series is the
``off shell" (namely keeping the ``source" $J$ nonzero) equation of state:

\begin{equation}
J^{A}=-T^{AX}\psi ^{X}-\frac{1}{3!}%
V^{AX_{2}X_{3}X_{4}}\left( \psi ^{X_{2}}\psi ^{X_{3}}\psi ^{X_{4}}+3\psi
^{X_{2}}\left \langle X_{3}X_{4}\right \rangle +\left \langle
X_{2}X_{3}X_{4}\right \rangle \right) \text{.}  \label{eqmotion}
\end{equation}%
On the right hand side the off shell expectation value $\left \langle \psi
\right \rangle $ is written simply as $\psi $. Due to the fact that the on -
shell ($J^{A}=0$) expectation value of odd number of fermionic variable
vanishes, the on - shell equation is trivially obeyed. The first nontrivial
equation, contain nonzero on - shell correlators within CQA is the gap
equation. It is the derivative of the equation of motion with respect to the
source:
\begin{eqnarray}
\frac{\delta }{\delta J^{B}}J^{A} &=&\delta ^{AB}=-T^{AX}\left \langle
BX\right \rangle -\frac{1}{2}V^{AXX_{3}X_{4}}\left \langle BX\right \rangle
\psi ^{X_{3}}\psi ^{X_{4}}-\frac{1}{2}V^{AXX_{3}X_{4}}\left \langle BX\right
\rangle \left \langle X_{3}X_{4}\right \rangle  \label{second} \\
&&+\frac{1}{2}V^{AX_{2}X_{3}X_{4}}\psi ^{X_{2}}\left \langle
BX_{3}X_{4}\right \rangle -\frac{1}{3!}V^{AX_{2}X_{3}X_{4}}\left \langle
BX_{2}X_{3}X_{4}\right \rangle \text{.}  \notag
\end{eqnarray}%
Furthermore the next successive derivative with respect to the source
results in,

\begin{eqnarray}
0 &=&\frac{\delta }{\delta J^{C}}\frac{\delta }{\delta J^{B}}%
J^{A}=-T^{AX}\left \langle CBX\right \rangle -\frac{1}{2}V^{AXX_{3}X_{4}}%
\left \langle CBX\right \rangle \psi ^{X_{3}}\psi
^{X_{4}}-V^{AXX_{3}X_{4}}\left \langle BX\right \rangle \left \langle
CX_{3}\right \rangle \psi ^{X_{4}}  \label{third} \\
&&-\frac{1}{2}V^{AXX_{3}X_{4}}\left \langle CBX\right \rangle \left \langle
X_{3}X_{4}\right \rangle -\frac{1}{2}V^{AXX_{3}X_{4}}\left \langle BX\right
\rangle \left \langle CX_{3}X_{4}\right \rangle +\frac{1}{2}%
V^{AX_{2}X_{3}X_{4}}\left \langle CX_{2}\right \rangle \left \langle
BX_{3}X_{4}\right \rangle  \notag \\
&&-\frac{1}{2}V^{AX_{2}X_{3}X_{4}}\psi ^{X_{2}}\left \langle
CBX_{3}X_{4}\right \rangle -\frac{1}{3!}V^{AX_{2}X_{3}X_{4}}\left \langle
CBX_{2}X_{3}X_{4}\right \rangle \text{.}  \notag
\end{eqnarray}%
The last term containing the five field correlator should be dropped within
the CQA. Further derivative with respect to $J_{D}$ will be required on
shell only. It reads,

\begin{multline}
-T^{AX}\left \langle DCBX\right \rangle -\frac{1}{2}\left \langle
X_{3}X_{4}\right \rangle V^{AXX_{3}X_{4}}\left \langle DCBX\right \rangle -%
\frac{1}{2}V^{AXX_{3}X_{4}}\left \langle BX\right \rangle \left \langle
DCX_{3}X_{4}\right \rangle \\
-\frac{1}{2}V^{AX_{2}X_{3}X_{4}}\left \langle DX_{2}\right \rangle \left
\langle CBX_{3}X_{4}\right \rangle +\frac{1}{2}V^{AX_{2}X_{3}X_{4}}\left%
\langle CX_{2}\right \rangle \left \langle DBX_{3}X_{4}\right \rangle
=V^{AX_{2}X_{3}X_{4}}\left \langle BX_{2}\right \rangle \left \langle
CX_{3}\right \rangle \left \langle DX_{4}\right \rangle ,  \label{fourth}
\end{multline}%
and importantly is linear in the quartic (four field) correlators. This will
be called in what follows the ``Bethe - Salpeter" (BS) despite important
differences with the original form in the factorization approach to bound
states and many - body theory\cite{albrecht1998ab,onida2002electronic}.

The second DS equation, Eq.(\ref{second}), on - shell, namely dropping the
odd number of fields correlators, can be written as,

\begin{equation}
\delta ^{AB}=-\left[ H^{-1}\right] ^{AX}\left \langle BX\right \rangle -%
\frac{1}{3!}V^{AX_{2}X_{3}X_{4}}\left \langle BX_{2}X_{3}X_{4}\right \rangle
\text{,}  \label{onshellgap}
\end{equation}%
where it is convenient to define a matrix,
\begin{equation}
\left[ H^{-1}\right] ^{AX}\equiv T^{AX}+\frac{1}{2}V^{AXX_{1}X_{2}}\left%
\langle X_{1}X_{2}\right \rangle \text{.}  \label{Hdef}
\end{equation}%
Multiplying from right by the matrix $H$, Eq.(\ref{Hdef}) becomes:
\begin{equation}
\delta ^{AB}=\left( T^{AX}+\frac{1}{2}V^{AXX_{1}X_{2}}\left \langle
X_{1}X_{2}\right \rangle \right) H^{XB}\text{.}  \label{deltaH}
\end{equation}%
Then Eq.(\ref{onshellgap}) takes a form that will be called the ``gap
equation" (again, despite obvious differences with the form in
the gaussian approximation):

\begin{equation}
\left \langle AB\right \rangle =-H^{BX_{1}}\left( \delta ^{X_{1}A}+\frac{1}{%
3!}V^{X_{1}X_{2}X_{3}X_{4}}\left \langle AX_{2}X_{3}X_{4}\right \rangle
\right) \text{.}  \label{gapeq}
\end{equation}%
A remarkable feature of CQA is that it does not require the covariant
corrections to the one body correlator. This is the most important finding
of the present work, since it allows a significant simplification of the
quartic scheme.

\subsection{Covariance of two field correlator.}

The full covariant two field correlator (or Green function) is $G_{\mathrm{%
full}}^{A_{1}A_{2}}=-\frac{\delta ^{2}F}{\delta J^{A_{1}}\delta J^{A_{2}}}=%
\frac{\delta ^{2}\left \langle \psi ^{A_{2}}\right \rangle }{\delta J^{A_{1}}%
}$. After first taking the derivative of the equation of motion, Eq.(\ref%
{eqmotion}), then taking $J^{X}=0$, one obtains:

\begin{equation}
\delta^{AB}=-T^{AX}\left \langle BX\right \rangle _{\mathrm{full}}-\frac{1}{2%
}V^{AXX_{3}X_{4}}\left \langle BX\right \rangle _{\mathrm{full}}\left
\langle X_{3}X_{4}\right \rangle -\frac{1}{3!}V^{AX_{2}X_{3}X_{4}}\frac{%
\delta}{\delta J^{B}}\left \langle X_{2}X_{3}X_{4}\right \rangle \text{.}
\label{fullcorreq}
\end{equation}
Multiplying this equation by $H$ defined in Eq.(\ref{Hdef}) as above one
obtains:

\begin{equation}
\left \langle AB\right \rangle _{\mathrm{full}}=-H^{BX_{1}}\left(%
\delta^{X_{1}A}+\frac{1}{3!}V^{X_{1}X_{2}X_{3}X_{4}}\frac{\delta}{\delta
J^{A}}\left \langle X_{2}X_{3}X_{4}\right \rangle \right).  \label{fullcorr}
\end{equation}
It involves the ``chain" that should be calculated by differentiating the
off - shell third DS equation (as in the cubic approximation developed in
refs. \onlinecite{Wang17,Cubic}):
\begin{equation}
\frac{\delta}{\delta J^{D}}\left \langle CBA\right \rangle =-\frac{1}{2}%
H^{AX_{1}}V^{X_{1}X_{2}X_{3}X_{4}}\left \{
\begin{array}{c}
2\left \langle BX_{2}\right \rangle \left \langle CX_{3}\right \rangle \left
\langle DX_{4}\right \rangle _{f}+\left \langle DX_{2}\right \rangle \left
\langle CBX_{3}X_{4}\right \rangle \\
+\left \langle BX_{2}\right \rangle \frac{\delta}{\delta J^{D}}\left \langle
CX_{3}X_{4}\right \rangle -\left \langle CX_{2}\right \rangle \frac{\delta}{%
\delta J^{D}}\left \langle BX_{3}X_{4}\right \rangle%
\end{array}%
\right \} \text{.}  \label{chain}
\end{equation}

Note that this equation is identical to the BS equation Eq.(\ref{fourth}),
so that, assuming that the solution is unique, one concludes that $\frac{%
\delta}{\delta J^{D}}\left \langle CBA\right \rangle =\left \langle
DCBA\right \rangle $ and consequently
\begin{equation}
\left \langle AB\right \rangle _{\mathrm{full}}=\left \langle AB\right
\rangle \text{.}  \label{noncorrection}
\end{equation}
Therefore the two field correlator\textbf{\ is covariant without the
correction terms. }This is highly nontrivial. While this is also the case in
CGA (for fermions only), in CCA the correction is non zero and crucial to
ensure Ward identities of continue symmetries. Here it is automatic. The same
cannot be proved for the full four field correlator $\left \langle
ABCD\right \rangle _{\mathrm{full}}$. It is most probably not equal to $%
\left \langle ABCD\right \rangle $. The question is rather academic, since,
as will be demonstrated below, the complexity of calculation grows as a high
power of the system's size.

Similar proof can be applied to scalar fields. This has been already used in
Section \ref{sec:hierarchy}. It is important to simplify the minimization
equations by choosing optimal linear combinations of the correlators. This
will be done next.

\subsection{A more economic linear combination of the chains: V-chains}

The interaction chain (or simply V - chain) is defined as a linear
combination,
\begin{equation}
\left \langle CD|AB\right \rangle =V^{ABXY}\left \langle CDXY\right \rangle .
\label{Vchaindef}
\end{equation}%
It can be shown by inspecting the equations that the quantity 
is antisymmetric under independent permutations $C\leftrightarrow D$ and $%
A\leftrightarrow B$. It is useful to consider the four index quantities like
the V - chain and $V^{ABXY}$ as matrices with antisymmetric pair of indices $%
\left( A,B\right) $ forming a ``super - vector" index.

Multiplying the BS equation by $V$ as a matrix, one obtains,

\begin{gather}
V^{BAY_{2}Y_{1}}\left \langle DCY_{2}Y_{1}\right \rangle +\frac{1}{2}%
V^{BAY_{2}Y_{1}}H^{Y_{1}X_{1}}\left( \left \langle Y_{2}X_{2}\right \rangle
\left \langle DC|X_{1}X_{2}\right \rangle -\left \langle CX_{2}\right
\rangle \left \langle DY_{2}|X_{1}X_{2}\right \rangle +\left \langle
DX_{2}\right \rangle \left \langle CY_{2}|X_{1}X_{2}\right \rangle \right)
\label{Vchain_deriv} \\
=-V^{BAY_{2}Y_{1}}H^{Y_{1}X_{1}}V^{X_{1}X_{2}X_{3}X_{4}}\left \langle
Y_{2}X_{2}\right \rangle \left \langle CX_{3}\right \rangle \left \langle
DX_{4}\right \rangle \text{,}  \notag
\end{gather}%
or, in terms of the V chains,
\begin{gather}
\left \langle DC|BA\right \rangle +\frac{1}{2}V^{BAY_{2}Y_{1}}H^{Y_{1}X_{1}}%
\left( \left \langle Y_{2}X_{2}\right \rangle \left \langle
DC|X_{1}X_{2}\right \rangle -\left \langle CX_{2}\right \rangle \left
\langle DY_{2}|X_{1}X_{2}\right \rangle +\left \langle DX_{2}\right \rangle
\left \langle CY_{2}|X_{1}X_{2}\right \rangle \right)  \label{BS_Vchain} \\
=-V^{BAY_{2}Y_{1}}V^{X_{1}X_{2}X_{3}X_{4}}H^{Y_{1}X_{1}}\left \langle
Y_{2}X_{2}\right \rangle \left \langle CX_{3}\right \rangle \left \langle
DX_{4}\right \rangle \text{.}  \notag
\end{gather}

Let us now apply this rather abstract formalism to a sufficiently general
charge conserving interacting electronic system (a ``many - body" problem).

\subsection{Electrons with pair - wise interactions}

\subsubsection{Action}

The Matsubara action of the general pair - wise interacting electron model
has the following (non Nambu) form\cite{NO} in terms of complex
Grassmannians:
\begin{equation}
\mathcal{A}\left[ \psi \right] =\psi _{a}^{\ast }T_{ab}\psi _{b}^{\cdot }+%
\frac{1}{2}\psi _{a}^{\ast }\psi _{a}^{\cdot }V_{ab}\psi _{b}^{\ast }\psi
_{b}^{\cdot }\text{.}  \label{manybody}
\end{equation}%
The interaction is of the density - density form (in most, but not all cases
originating from the Coulomb repulsion) and thus $V_{ab}=V_{ba}$. The
electric charge conservation is explicit here (number of $\psi $ and $\psi
^{\ast }$ is equal in each term). To relate the hopping term coefficient and
the interaction potential to the Nambu action of the previous subsection,
let us split Eq.(\ref{Matsubara_action}), into the charge (Nambu) index $%
A=\ast ,\cdot $, and the rest:

\begin{equation}
\mathcal{A}=\frac{1}{2}\psi _{a}^{A}T_{ab}^{AB}\psi _{b}^{B}+\frac{1}{4!}%
V_{abcd}^{ABCD}\psi _{a}^{A}\psi _{b}^{B}\psi _{c}^{C}\psi _{d}^{D}\text{.}
\label{Nambuform}
\end{equation}%
The result is:
\begin{eqnarray}
T_{ab}^{AB} &=&\delta ^{A\ast }\delta ^{B\cdot }T_{ab}-\delta ^{A\cdot
}\delta ^{B\ast }T_{ba}\text{;}  \label{TV} \\
V_{y_{1}y_{2}y_{3}y_{4}}^{Y_{1}Y_{2}Y_{3}Y_{4}} &=&\delta
_{y_{3}y_{4}}V_{y_{3}y_{1}}\delta _{y_{1}y_{2}}\left( \delta ^{\ast
Y_{4}}\delta ^{\cdot Y_{3}}-\delta ^{\cdot Y_{4}}\delta ^{\ast Y_{3}}\right)
\left( \delta ^{\ast Y_{2}}\delta ^{\cdot Y_{1}}-\delta ^{\cdot Y_{2}}\delta
^{\ast Y_{1}}\right)  \notag \\
&&-\delta _{y_{2}y_{4}}V_{y_{2}y_{1}}\delta _{y_{1}y_{3}}\left( \delta
^{\ast Y_{4}}\delta ^{\cdot Y_{2}}-\delta ^{\cdot Y_{4}}\delta ^{\ast
Y_{2}}\right) \left( \delta ^{\ast Y_{3}}\delta ^{\cdot Y_{1}}-\delta
^{\cdot Y_{3}}\delta ^{\ast Y_{1}}\right)  \notag \\
&&-\delta _{y_{1}y_{4}}V_{y_{3}y_{1}}\delta _{y_{3}y_{2}}\left( \delta
^{\ast Y_{4}}\delta ^{\cdot Y_{1}}-\delta ^{\cdot Y_{4}}\delta ^{\ast
Y_{1}}\right) \left( \delta ^{\ast Y_{2}}\delta ^{\cdot Y_{3}}-\delta
^{\cdot Y_{2}}\delta ^{\ast Y_{3}}\right) \text{.}  \notag
\end{eqnarray}%
This model is now amenable to the CQA scheme described in the previous
subsections.

\subsubsection{The minimization equations}

Assuming that the charge symmetry is not spontaneously broken (no
superconductivity), which leads to two and four field correlators without
charge conserving zero, the following notations for Green function and the
V - chains will
be used:
\begin{eqnarray}
\left \langle _{ab}^{\ast \cdot}\right \rangle & = & -\left \langle
_{ba}^{\cdot \ast}\right \rangle =G_{ab};H_{ab}^{\ast \cdot}=-H_{ba}^{\cdot
\ast}=H_{ab};  \label{Vchainmoddef} \\
\left \langle _{ab}^{\ast \cdot}|_{cd}^{\ast \cdot}\right \rangle & = &
D_{abcd};\text{ \  \ }\left \langle _{ab}^{\ast \ast}|_{cd}^{\ast \ast}\right
\rangle =R_{abcd};\text{ \  \  \ }\left \langle _{ab}^{\cdot
\cdot}|_{cd}^{\cdot \cdot}\right \rangle =C_{abcd}  \notag
\end{eqnarray}
Here $D$ is the diffuson chain, while $C$ is Cooperon chain. It turns out,
at least for a local interaction, that $R_{abcd}$ is related to $C_{abcd}$
by complex conjugation with Matsubara times reflected (see precise relation
of Fourier components below). Let us now turn to the minimization equations.

The $\cdot \cdot$ component of the definition of $H^{AB}$ Eq.(\ref{deltaH})
is now
\begin{equation}
\delta_{ab}=\left(-T_{xa}+\left(-\delta_{ax}V_{ya}G_{yy}+V_{xa}G_{ax}\right)%
\right)H_{xb}\text{.}  \label{Hmodel}
\end{equation}
For nonsuperconducting states one obtains the gap equation, Eq.(\ref{gapeq}%
), for the charge components $\ast \cdot$ in the form:

\begin{equation}
G_{ab}=H_{x_{1}b}\left(\delta_{x_{1}a}+\frac{1}{6}%
\left(D_{x_{1}x_{2}ax_{2}}+R_{x_{1}x_{2}ax_{2}}\right)\right)\text{.}
\label{gapmodel}
\end{equation}
Obviously the last term makes a profound difference compared to various gap
equations encountered in Hartree - Fock type methods. It couples the two
field correlator to a particular linear combination of the four field
correlator that enters the quartic or BS like equation. The BS equation
however is much more involved.

Since the BS equations are linear in the chains variables,$C$, $R$ and $T$,
let us write it using a (double index) matrix form. The pair of the diffuson
equation is,

\begin{eqnarray}
D_{abcd}+L_{abcd} &=&U_{abcd}\text{;}  \label{chaindef} \\
C_{abcd}+M_{abcd} &=&W_{abcd}\text{,}  \notag
\end{eqnarray}%
where the homogeneous terms are,
\begin{eqnarray}
L_{abcd} &=&-\frac{V_{ab}}{2}\left(
\begin{array}{c}
H_{bx_{1}}G_{x_{2}a}D_{x_{2}x_{1}cd}-H_{bx_{1}}G_{x_{2}d}D_{x_{2}x_{1}ca}-H_{x_{1}a}G_{cx_{2}}D_{x_{1}x_{2}bd}
\\
+H_{x_{1}a}G_{bx_{2}}D_{x_{1}x_{2}cd}-H_{bx_{1}}G_{cx_{2}}C_{x_{1}x_{2}ad}-H_{x_{1}a}G_{x_{2}d}R_{x_{1}x_{2}bc}%
\end{array}%
\right)  \label{homogeneous} \\
&&+\frac{\delta _{ab}V_{ya}}{2}\left(
\begin{array}{c}
H_{yx_{1}}G_{x_{2}y}D_{x_{2}x_{1}cd}-H_{x_{1}y}G_{yx_{2}}D_{x_{1}x_{2}cd}+H_{yx_{1}}G_{x_{2}d}D_{x_{2}x_{1}cy}
\\
+H_{x_{1}y}G_{cx_{2}}D_{x_{1}x_{2}yd}+H_{yx_{1}}G_{cx_{2}}C_{x_{1}x_{2}yd}+H_{x_{1}y}G_{x_{2}d}R_{x_{1}x_{2}yc}%
\end{array}%
\right) \text{;}  \notag \\
M_{abcd} &=&\frac{V_{ba}}{2}\left(
H_{ax_{1}}G_{bx_{2}}C_{x_{1}x_{2}cd}-H_{bx_{1}}G_{ax_{2}}C_{x_{1}x_{2}cd}+H_{bx_{2}}G_{x_{1}c}D_{x_{1}x_{2}ad}\right)
\notag \\
&&-\frac{V_{ba}}{2}\left(
H_{ax_{2}}G_{x_{1}c}D_{x_{1}x_{2}bd}+H_{bx_{2}}G_{x_{1}d}D_{x_{1}x_{2}ac}-H_{ax_{2}}G_{x_{1}d}D_{x_{1}x_{2}bc}\right)
\text{,}  \notag
\end{eqnarray}%
and the inhomogeneous are

\begin{eqnarray}
U_{abcd} &=&V_{x_{1}x_{2}}\left \{
\begin{array}{c}
V_{ab}\left( -G_{cx_{1}}G_{x_{2}d}\left(
H_{bx_{2}}G_{x_{1}a}+H_{x_{1}a}G_{bx_{2}}\right) +G_{cx_{2}}G_{x_{2}d}\left(
H_{bx_{1}}G_{x_{1}a}+H_{x_{1}a}G_{bx_{1}}\right) \right) \\
+\delta _{ab}V_{ya}\left( G_{cx_{1}}G_{x_{2}d}\left(
H_{yx_{2}}G_{x_{1}y}+H_{x_{1}y}G_{yx_{2}}\right) -G_{cx_{2}}G_{x_{2}d}\left(
H_{yx_{1}}G_{x_{1}y}+H_{x_{1}y}G_{yx_{1}}\right) \right)%
\end{array}%
\right \} \text{;}  \label{vector} \\
W_{abcd} &=&V_{ab}V_{x_{1}x_{2}}\left \{
-H_{bx_{1}}G_{ax_{2}}G_{x_{2}c}G_{x_{1}d}+H_{bx_{1}}G_{ax_{2}}G_{x_{1}c}G_{x_{2}d}+H_{ax_{1}}G_{bx_{2}}G_{x_{2}c}G_{x_{1}d}-H_{ax_{1}}G_{bx_{2}}G_{x_{1}c}G_{x_{2}d}\right \}
\text{.}  \notag
\end{eqnarray}

This is quite cumbersome, but translation invariance in time and space (for
homogeneous phases, antiferromagnets for example have less translation
symmetry) can be used to make simplifications. In addition, we will limit
ourselves in this paper to Hubbard model in $D=1,2$ dimensions. Locality of
interactions also simplifies significantly the minimization equations.
Therefore we specify the discussion to local interactions from now on. Exact
solution exists for sufficiently small $N$ in the case of local interaction
- the Hubbard model. We therefore apply CQA to the case of Hubbard model and
compare it to the exact diagonalization\cite{ED} (ED) in the case of 1D with
small finite $N$ and to Monte Carlo simulations at half filling in 2D.

$\ $

\section{The CQA approximation in the Hubbard model}

\label{sec:Hubbardformula}

\subsection{The single band Hubbard model in $D$ dimensions and its symmetries}

\subsubsection{Hamiltonian and the Matsubara action}

The single band Hubbard model is defined on the $D$ dimensional hypercubic
lattice. The tunneling amplitude to the neighboring site in any direction $%
i=1,...,D$ is denoted in literature by $t$. We choose it to be the unit of
energy $t=1$. Similarly the lattice spacing sets the unit of length and $%
\hbar=1$. The Hamiltonian is:

\begin{equation}
H=\sum \nolimits_{r_{1,...},r_{D}}\left \{ -\sum \nolimits_{i}\left(
a_{r}^{A\dagger }a_{r+\widehat{i}}^{A}+h.c.\right) -\mu
n_{r}+Un_{r}^{\upharpoonleft }n_{r}^{\downarrow }\right \} \text{.}
\label{HubbardH}
\end{equation}%
The chemical potential $\mu $ and the on - site repulsion energy $U$ (given
in units of the hopping energy). The spin index takes two values $%
A,B=\uparrow ,\downarrow $. The hopping direction is denoted by $\widehat{i}$%
. The density and its spin components are $n_{r}=n_{r}^{\upharpoonleft
}+n_{r}^{\downarrow }$ and $n_{r}^{A}\equiv a_{r}^{A\dagger }a_{r}^{A}$. It
is well known that at half filling $\mu =\frac{U}{2}$ due to particle - hole
symmetry, and we concentrate mostly on this case.

The simplest discretized Matsubara action is\cite{NO},
\begin{eqnarray}
\mathcal{A} &=&\tau \sum \nolimits_{t,r}\left \{ \frac{1}{\tau }\left( \psi
_{t+1,r}^{A\ast }\psi _{t,r}^{A}-\psi _{t,r}^{A\ast }\psi _{t,r}^{A\ast
}\right) \ -\frac{1}{2}\sum \nolimits_{i}\left( \psi _{t,r}^{A\dagger }\psi
_{t,r+\widehat{i}}^{A}+\psi _{t,x}^{A\dagger }\psi _{t,r-\widehat{i}%
}^{A}\right) \right. \text{,}  \label{Liuaction} \\
&&\left. -\mu n_{t,r}-U\psi _{t,r}^{\upharpoonleft \ast }\psi
_{t,r}^{\downarrow \ast }\psi _{t,r}^{\upharpoonleft }\psi
_{t,r}^{\downarrow }\right \}   \notag
\end{eqnarray}%
where $n_{t,r}\equiv \psi _{t,r}^{X\dagger }\psi _{t,r}^{X}$ and the
Matsubara time is on the
circle of circumference $1/T$, where $T$ is temperature. One discretizes the
path integral into $M$ segments with the Matsubara time step $\tau =\left(
TM\right) ^{-1}$, so that $t$ is an integer variable\cite{NO} taking values $t=0,..,M-1$.

\subsubsection{Symmetry considerations}

Although the general equations can be written by substituting the Hubbard
hopping and interaction into general formulas of the previous section, here
we concentrate on a simpler particular case of the paramagnetic phase in
which the ground state has full spin rotation $SU\left( 2\right) $ symmetry,
\begin{equation}
\psi _{t,r}^{A}\rightarrow U^{AB}\psi _{t,r}^{B}\text{,}  \label{symmetry}
\end{equation}%
for any two dimensional unitary spin matrix $U$. The translation on the
periodic lattice symmetry is also not broken in low dimension $D\leq 2$ (as
it would be in antiferromagnet with long range order in $D=3$). We therefore
do not consider $D=3$. Generally for systems in $D\leq 2$ with continuous symmetry
 fluctuations (quantum and thermal) destroy broken phases\cite%
{amit2005field,chaikin1995principles}, although previously attempted
variational approaches like the CGA (extending Hartree - Fock) and even CCA
at large coupling can start from a ``broken" phase solution of the
minimization equations sometimes give a better result upon symmetrization.
This was the main topic of previous paper\cite{symmetrization}. It turns out
that CQA allows only paramagnetic solutions of the minimization equation of
the half filled Hubbard model in low dimension, namely antiferromagnetic and
ferromagnetic solutions do not exist, consistent with Mermin - Wagner
theorem, in contrast to Hartree - Fock or Gauss approximation where the
symmetry breaking (spurious) solution exists even in low dimension $D\leq 2$%
. For paramagnetic solution covariance under the spin rotation $SU\left(
2\right) $ group immediately results in:
\begin{eqnarray}
G_{ab}^{AB} &=&\delta ^{AB}G_{ab};  \label{paramagnet} \\
H_{ab}^{AB} &=&\delta ^{AB}H_{ab}\text{,}  \notag
\end{eqnarray}%
where $A,B$ are spin indices here now, and $a,b$ are the space time coordinates.

\subsection{Simplification of the minimization equations due to the
interaction locality and the spin rotation invariance.}

The gap equation, Eq.(\ref{gapmodel}) now is simplified into

\begin{equation}
\delta^{AB}G_{ab}=\delta^{AB}H_{ab}+\frac{1}{6}H_{xb}%
\left(D_{xyay}^{BXAX}+R_{xyay}^{BXAX}\right)\text{.}  \label{gapeq1}
\end{equation}
The homogeneous terms of quartic equations are much more involved:

\begin{eqnarray}
L_{abcd}^{ABCD} &=&\frac{\tau U}{2}\delta _{ab}\left \{ -\left(
\begin{array}{c}
H_{ax_{1}}G_{x_{2}a}D_{x_{2}x_{1}cd}^{ABCD}-H_{ax_{1}}G_{x_{2}d}D_{x_{2}x_{1}ca}^{DBCA}-H_{x_{1}a}G_{cx_{2}}D_{x_{1}x_{2}ad}^{ACBD}
\\
+H_{x_{1}a}G_{ax_{2}}D_{x_{1}x_{2}cd}^{ABCD}-H_{ax_{1}}G_{cx_{2}}C_{x_{1}x_{2}ad}^{BCAD}-H_{x_{1}a}G_{x_{2}d}R_{x_{1}x_{2}ac}^{ADBC}%
\end{array}%
\right) \right.  \label{bs1} \\
&&\left. +\delta _{AB}\left(
\begin{array}{c}
H_{yx_{1}}G_{x_{2}a}D_{x_{2}x_{1}cd}^{YYCD}-H_{x_{1}a}G_{ax_{2}}D_{x_{1}x_{2}cd}^{YYCD}+H_{ax_{1}}G_{x_{2}d}D_{x_{2}x_{1}ca}^{DYCY}
\\
+H_{x_{1}a}G_{cx_{2}}D_{x_{1}x_{2}ad}^{YCYD}+H_{ax_{1}}G_{cx_{2}}C_{x_{1}x_{2}ad}^{YCYD}+H_{x_{1}a}G_{x_{2}d}R_{x_{1}x_{2}ac}^{YDYC}%
\end{array}%
\right) \right \} ; \\
M_{abcd}^{ABCD} &=&\frac{\tau U}{2}\delta _{ab}\left(
\begin{array}{c}
H_{ax_{1}}G_{bx_{2}}C_{x_{1}x_{2}cd}^{ABCD}-H_{ax_{1}}G_{ax_{2}}C_{x_{1}x_{2}cd}^{BACD}+H_{ax_{2}}G_{x_{1}c}D_{x_{1}x_{2}ad}^{CBAD}
\\
-H_{ax_{2}}G_{x_{1}c}D_{x_{1}x_{2}ad}^{CABD}-H_{ax_{2}}G_{x_{1}d}D_{x_{1}x_{2}ac}^{DBAC}+H_{ax_{2}}G_{x_{1}d}D_{x_{1}x_{2}ac}^{DABC}%
\end{array}%
\right) \text{,}  \notag
\end{eqnarray}%
while the inhomogeneous terms are
\begin{eqnarray}
U_{abcd}^{ABCD} &=&\tau ^{2}U^{2}\left( \delta ^{AB}\delta ^{CD}-\delta
^{AC}\delta ^{BD}\right) \delta _{ab}G_{cx}G_{xd}\left(
H_{ax}G_{xa}+H_{xa}G_{ax}\right) \text{;}  \label{bsvec} \\
W_{abcd}^{ABCD} &=&-2\tau ^{2}U^{2}\left( \delta ^{AC}\delta ^{BD}-\delta
^{AD}\delta ^{BC}\right) \delta _{ab}H_{ax}G_{ax}G_{xc}G_{xd}\text{.}  \notag
\end{eqnarray}

The paramagnetic Ansatz for diffusons can be inferred from the inhomogeneous
parts of the BS equations (which represent the leading order in perturbation
theory), Eq.(\ref{bsvec}):
\begin{equation}
D_{abcd}^{ABCD}=\delta_{ab}\left(\delta_{CD}^{AB}D_{acd}^{1}+%
\delta_{BD}^{AC}D_{acd}^{2}\right)\text{.}  \label{spindif}
\end{equation}
For cooperons the analogous dependence is:

\begin{equation}
C_{abcd}^{ABCD}=\delta_{ab}\left(\delta_{BD}^{AC}-\delta_{BC}^{AD}%
\right)C_{acd}\text{,}\   \label{spincoop}
\end{equation}
and same for $R$. Due to locality and instantaneity of the interaction
reflected in Eq.(\ref{bsvec}) one needs only the coincident space/time
points V - chains, $D_{abc}^{1,2}=D_{aabc}^{1,2}$.

Substituting the spin Ansatz in\ the gap equation, one obtains:

\begin{eqnarray}
G_{ab} & = & H_{ab}+\frac{1}{6}H_{xb}\left(D_{xax}^{1}+2D_{xax}^{2}+R_{xax}%
\right);  \label{gapHub} \\
\delta_{ab} & = & -\left(T_{xa}+U\delta_{ax}G_{aa}\right)H_{xb}\text{.}
\notag
\end{eqnarray}
Subsequently the diffuson chain equations take a form,

\begin{eqnarray}
D_{acd}^{1}+\frac{\tau U}{2}\left \{
\begin{array}{c}
H_{ax}G_{xd}\left(D_{xca}^{1}+D_{xca}^{2}\right)+H_{xa}G_{cx}%
\left(D_{xad}^{1}+D_{xad}^{2}\right) \\
-\left(H_{xa}G_{ax}+H_{ax}G_{xa}\right)\left(D_{xcd}^{1}+D_{xcd}^{2}\right)%
\end{array}%
\right \} & = & 0\text{;} \\
D_{acd}^{2}+\frac{\tau U}{2}\left \{
\begin{array}{c}
\left(H_{xa}G_{ax}+H_{ax}G_{xa}\right)D_{xcd}^{2}-H_{ax}G_{xd}D_{xca}^{1} \\
-H_{xa}G_{cx}D_{xad}^{1}+H_{ax}G_{cx}C_{xad}+H_{xa}G_{xd}R_{xac}%
\end{array}%
\right \} & = &
-\tau^{2}U^{2}\left(H_{ax}G_{xa}+H_{xa}G_{ax}\right)G_{cx}G_{xd}\text{,}
\notag
\end{eqnarray}
while the cooperon equations are,
\begin{eqnarray}
C_{acd}+\frac{\tau U}{2}\left \{
\begin{array}{c}
2H_{ax}G_{ax}C_{xcd}+H_{ax}G_{xc}\left(D_{xad}^{2}-D_{xad}^{1}\right) \\
+H_{ax}G_{xd}\left(D_{xac}^{2}-D_{xac}^{1}\right)%
\end{array}%
\right \} & = & -2\tau^{2}U^{2}H_{ax}G_{ax}G_{xc}G_{xd}\text{;} \\
R_{acd}+\frac{\tau U}{2}\left \{
\begin{array}{c}
2H_{xa}G_{xa}R_{xcd}+H_{xa}G_{cx}\left(D_{xda}^{2}-D_{xda}^{1}\right) \\
+H_{xa}G_{dx}\left(D_{xca}^{2}-D_{xca}^{1}\right)%
\end{array}%
\right \} & = & -2\tau^{2}U^{2}H_{xa}G_{xa}G_{cx}G_{dx}\text{.}  \notag
\end{eqnarray}
Further simplification occurs when the translation invariance is utilized
below.

\subsection{Translation invariance, DSE in Fourier form}

The model is constructed on the lattice with periodic boundary conditions $N$
in each direction, to keep notations as simple as possible, the square
lattice is assumed with lattice spacing defining the unit of length. The
points therefore are $r_{i}=1,..N$, $i=1,..,D$ (dimensionality). At
temperature $T$ the Matsubara (Euclidean) time is also discretized $%
t=0,...M-1$ in the range $0<\tau t\leq 1/T$ and $\psi _{t,r}^{A}$ is
antiperiodic in $t$\cite{NO}.

The discrete space - time index $a$ will be eventually substituted by
integer valued wave number $k$ and the Matsubara frequency $n$:
\begin{equation}
\psi _{a}^{A\ast }=\sqrt{\frac{T}{N^{D}}}\sum%
\nolimits_{k_{1},...k_{D}=1}^{N}\sum \nolimits_{n=1}^{M}\exp \left[ -2\pi
i\left( \frac{\left( n+1/2\right) t}{M}+\frac{k_{i}r_{i}}{N}\right) \right]
\psi _{\alpha }^{A\ast }\text{,}  \label{FTdef}
\end{equation}%
where $\alpha =\left \{ n,k_{1},...,k_{D}\right \} $ enumerates the space -
time components of the frequency - quasi-momentum basis. The translation
invariance (the energy and the momentum conservation) leads to the following
Fourier transforms for the correlators:
\begin{equation}
G_{ab}=\frac{T}{N^{D}}\sum \nolimits_{\alpha }\exp \left[ i\left( b-a\right)
\cdot \alpha \right] g_{\alpha }\text{,}  \label{Gft}
\end{equation}%
where $\mathbf{\alpha }=2\pi \left \{ \frac{n+1/2}{M},\frac{k_{1}}{N},...,%
\frac{k_{D}}{N}\right \} $. The Fourier transform of tunneling amplitude is,
\begin{equation}
T_{ab}=\frac{\tau }{MN^{D}}\sum \nolimits_{\alpha }\exp \left[ i\left(
a-b\right) \cdot \alpha \right] t_{\mathbf{\alpha }}\text{,}  \label{Tft}
\end{equation}%
and the same transformation for $H$. In the Hubbard model of the simplest
action given by Eq.(\ref{Liuaction}), it takes the following form:%
\begin{eqnarray}
t_{n\mathbf{k}} &=&i\omega _{n}-2\sum \nolimits_{i}\cos \left[ \frac{2\pi }{N%
}k_{i}\right] ;  \label{tNO} \\
i\omega _{n} &=&TM\left( \exp \left[ i\frac{2\pi }{M}\left( n+1/2\right) %
\right] -1\right) \text{.}  \notag
\end{eqnarray}%
Note that the frequency part is periodic: $\omega _{n}=\omega _{n+M\text{. }%
} $For $n/M<<1,$
\begin{equation}
\omega _{n}\approx \pi T\left( 2n+1\right) \equiv \omega _{n}^{M},
\label{Matsubaradef}
\end{equation}
(Matsubara frequency). Close to $M$, $1-n/M<<1$, one has $\omega _{n}\approx
\pi T\left( 2n-2M+1\right) $. Therefore for $M\rightarrow \infty $ one
recovers the Matsubara frequency for both positive and negative $n$.

The chain functions have the following Fourier transforms (different for
diffusons and cooperons due to charges involved): 

\begin{align}
D_{abc}^{1,2}& =\frac{1}{\tau \left( MN^{D}\right) ^{2}}\sum
\nolimits_{\beta \gamma }\exp \left[ i\left( \left( a-b\right) \cdot \beta
+\left( c-a\right) \cdot \gamma \right) \right] d_{\beta \gamma }^{1,2};
\notag \\
R_{abc}& =\frac{1}{\tau \left( MN^{D}\right) ^{2}}\sum \nolimits_{\beta
\gamma }\exp \left[ i\left( \left( a-b\right) \cdot \beta +\left( a-c\right)
\cdot \gamma \right) \right] r_{\beta \gamma };  \label{FTchain} \\
C_{abc}& =\frac{1}{\tau \left( MN^{D}\right) ^{2}}\sum \nolimits_{\beta
\gamma }\exp \left[ i\left( \left( b-a\right) \beta +\left( c-a\right)
\gamma \right) \right] c_{\beta \gamma }\text{.}  \notag
\end{align}%
The gap equation becomes,
\begin{equation}
g_{\omega }=h_{\omega }\left( 1+\frac{T}{6N^{D}}\sum_{\lambda }\left(
r_{\omega \lambda }+d_{\omega \lambda }^{1}+2d_{\omega \lambda }^{2}\right)
\right) \text{,}  \label{gap}
\end{equation}%
where
\begin{equation}
h_{\omega }=-\left( t_{\omega }+\frac{UT}{N^{D}}\sum_{\lambda }g_{\lambda
}\right) ^{-1}\text{.}  \label{h_eq}
\end{equation}%
The set of the BS equations takes the following form:
\begin{eqnarray}
d_{\beta \gamma }^{1}+\frac{UT}{2N^{D}}\sum_{\lambda }\left \{
\begin{array}{c}
h_{\gamma -\beta +\lambda }g_{\gamma }\left( d_{\beta \lambda }^{1}+d_{\beta
\lambda }^{2}\right) +h_{\beta -\gamma +\lambda }g_{\beta }\left( d_{\lambda
\gamma }^{1}+d_{\lambda \gamma }^{2}\right) \\
-h_{\nu }\left( g_{\gamma -\beta +\nu }+g_{\beta -\gamma +\nu }\right)
\left( d_{\beta \gamma }^{1}+d_{\beta \gamma }^{2}\right)%
\end{array}%
\right \} &=&0;  \label{chaineqFT} \\
d_{\beta \gamma }^{2}+\frac{UT}{2N^{D}}\sum_{\lambda }\left \{
\begin{array}{c}
h_{\lambda }\left( g_{\lambda +\gamma -\beta }+g_{\beta +\lambda -\gamma
}\right) d_{\beta \gamma }^{2}-h_{\beta -\gamma +\lambda }g_{\beta
}d_{\lambda \gamma }^{1} \\
-h_{\gamma -\beta +\lambda }g_{\gamma }d_{\beta \lambda }^{1}+h_{\gamma
-\beta +\lambda }g_{\beta }c_{\lambda \gamma }+h_{\beta -\gamma +\lambda
}g_{\gamma }r_{\lambda \beta }%
\end{array}%
\right \} &=&-\frac{U^{2}T}{N^{D}}\sum_{\lambda }h_{\lambda }g_{\beta
}g_{\gamma }\left( g_{\beta -\gamma +\lambda }+g_{\gamma -\beta +\lambda
}\right) ;  \notag \\
c_{\beta \gamma }+\frac{UT}{N^{D}}\sum_{\lambda }h_{\lambda }g_{\beta
+\gamma -\lambda }c_{\beta \gamma }+\frac{UT}{2N}\sum_{\lambda }h_{\beta
+\gamma -\lambda }\left \{ g_{\beta }\left( d_{\lambda \gamma
}^{2}-d_{\lambda \gamma }^{1}\right) +g_{\gamma }\left( d_{\lambda \beta
}^{2}-d_{\lambda \beta }^{1}\right) \right \} &=&-\frac{2U^{2}T}{N^{D}}%
\sum_{\lambda }h_{\lambda }g_{\beta }g_{\gamma }g_{\beta +\gamma -\lambda };
\notag \\
r_{\beta \gamma }+\frac{UT}{N^{D}}\sum_{\lambda }h_{\lambda }g_{\beta
+\gamma -\lambda }r_{\beta \gamma }+\frac{UT}{2N}h_{\beta +\gamma -\lambda
}\left \{ g_{\beta }\left( d_{\gamma \lambda }^{2}-d_{\gamma \lambda
}^{1}\right) +g_{\gamma }\left( d_{\beta \lambda }^{2}-d_{\beta \lambda
}^{1}\right) \right \} &=&-\frac{2U^{2}T}{N^{D}}\sum_{\lambda }h_{\lambda
}g_{\beta }g_{\gamma }g_{\beta +\gamma -\lambda }\text{.}  \notag
\end{eqnarray}

The large set of generally nonlinear equations are solved numerically by
iteration described in the next section.

\section{Implementation and comparison with exact results for one and two
dimensional Hubbard model}

\label{sec:implement}

In this section computational issues of the calculation are described.

\subsection{The frequency cutoff}

The summations over the fermion Matsubara frequencies in Eq.(\ref{h_eq}), are
tricky. For far away from the time translation invariant frequency the
asymptotics of the chain function is $1/\omega _{n}^{2}$, so that all the
other summations over frequencies converge fast. The density summation
however converges slowly for the simplest action. A way to speed up the
calculation is to replace the frequency in tunneling amplitude Eq.(\ref{tNO}%
) by the Matsubara frequency
\begin{equation}
\omega _{n}=\pi T\left( 2n+1\right)  \label{perfectaction}
\end{equation}%
and introduce a frequency cutoff $M_{^{\Omega}}$, out of which
($ n \ge M_{^{\Omega }}$ or $n < -M_{^\Omega}$)
 $g\left( k,n\right) \approx \frac{-1}{i \omega_{n}}$, and the chain
functions are even smaller and assumed to be zero if one of Matsubara frequency of the chain
function is out of the cutoff.

There are infinite summations
over fermion Matsubara frequency $T\sum_{\lambda }g_{\lambda }$. However the
summation of the Matsubara frequency,$T\sum_{n=-\infty }^{\infty }g\left(
n,k\right) $ seems diverge. Actually the summation from the functional
approach, there is a damping factor $T\sum_{n=-\infty }^{\infty }g\left(
n,k\right) e^{-i\omega _{n}\eta }$ where $\eta =\beta /M$ and $\eta
\rightarrow +0$ when $M\rightarrow \infty $. Using the cutoff assumption, $%
g\left( n,k\right) =-\frac{1}{i\omega _{n}}$, for $ n \ge M_{^{\Omega }}$
 or $n < -M_{^\Omega}$:
\begin{eqnarray}
n_{k} &=&\lim_{\eta \downarrow 0}T\sum_{n=-\infty }^{\infty }g\left(
n,k\right) e^{-i\omega _{n}\eta }  \\
&\approx &T\sum_{n=-M_{^{\Omega }}}^{M_{^{\Omega }}-1}\left[ g\left(
n,k\right) +\frac{1}{i\omega _{n}}\right] +\lim_{\eta \downarrow
0}T\sum_{n=-\infty }^{\infty }\frac{-1}{i\omega _{n}}e^{-i\omega _{n}\eta }=%
\frac{1}{2}+T\sum_{n=-M_{^{\Omega }}}^{M_{^{\Omega }}-1}g\left( n,k\right) .
\notag
\end{eqnarray}%
With this definition the sum over the correlator is replaced by%
\begin{equation}
n_{k}\approx \frac{1}{2}+T\sum_{n=-M_{^{\Omega }}}^{M_{^{\Omega }}-1}g_{n%
\mathbf{k}}\text{.}  \label{addition}
\end{equation}%
$M_{^{\Omega }}$ will be large enough in order to get a convergent result $%
g_{n\mathbf{k}}$ for small $n$. All results presented in this paper are
calculated at $M_{^{\Omega }}=128$ and checked that it is indeed a
convergent result.


Given a $M_{^{\Omega }}$, we solve the quartic minimization equations
iteratively, as described below.

\subsection{The iterative solution of the minimization equations}

An effective iteration scheme for solution of the chain equations is
obtained when several simple (but often large) homogeneous terms are
combined with the left hand side of Eqs.(\ref{chaineqFT}). For example the $%
d_{\beta \gamma }^{2}$ equation can be rearranged as (Einstein summation
over $\lambda $ assumed),%
\begin{equation}
d_{\beta \gamma }^{2}\left \{ 1+\frac{UT}{2N}h_{\lambda }\left( g_{\lambda
+\gamma -\beta }+g_{\beta +\lambda -\gamma }\right) \right \} =\mathrm{RHS},
\end{equation}%
where

\begin{eqnarray}
\mathrm{RHS} &=&\frac{UT}{2N}g_{\beta }h_{\beta -\gamma +\lambda }d_{\lambda
\gamma }^{1}-\frac{UT}{2N}\left( -h_{\gamma -\beta +\lambda }g_{\gamma
}d_{\beta \lambda }^{1}+h_{\gamma -\beta +\lambda }g_{\beta }c_{\lambda
\gamma }+h_{\beta -\gamma +\lambda }g_{\gamma }r_{\lambda \beta }\right)
\notag \\
&&-\frac{U^{2}T}{N}h_{\lambda }g_{\beta }g_{\gamma }\left( g_{\beta -\gamma
+\lambda }+g_{\gamma -\beta +\lambda }\right) \text{,}
\end{eqnarray}%
and iteration formula is%
\begin{equation}
d_{\beta \gamma }^{2}=\frac{\mathrm{RHS}}{1+\frac{UT}{2N}h_{\lambda }\left(
g_{\lambda +\gamma -\beta }+g_{\beta +\lambda -\gamma }\right) }\text{.}
\end{equation}

The summation was accelerated by using fast Fourier transform and parallel
computing during the iteration process which will be elaborated in Fig. \ref%
{fig:flowchart} in Section \ref{sec:complexity}.

\subsection{Comparison with exact diagonalization(1D)}

\begin{figure}[htbp]
\centering
\includegraphics[width=\linewidth]{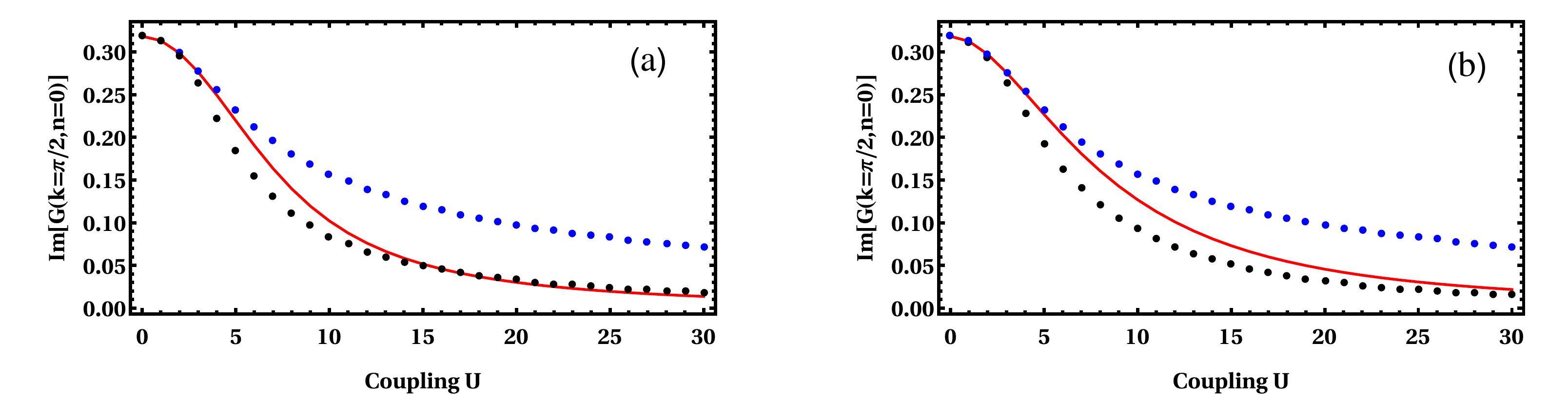}
\caption{Comparison of GW, covariant quartic approximation and exact result
for imaginary part of the Green function of 1D Hubbard model for (a)$T=1$,$%
\protect \delta \protect \mu=0$ and (b)$T=1$,$\protect \delta \protect \mu=0.2U$%
. The red line is the exact correlator, while blue and black dots are GW and
CQA respectively.}
\label{fig:gvsU}
\end{figure}

The one dimensional case is considered for simplicity and availability of
exact results utilizing the exact diagonalization\cite{ED} for small values
of $N$. We exactly calculate Green's function for $N=4$ and $T=1$
 for two values of $\delta \mu=\mu-\frac{U}{2}$ which are compared
with CQA results. $\delta \mu=0$ and $\delta \mu=0.2U$ are plotted in Fig.%
\ref{fig:gvsU}(a)and Fig.\ref{fig:gvsU}(b) respectively. By using fast
Fourier transform and parallel computing at The High-performance Computing
Platform of Peking University (one node only) with 32 cores during the
iteration process to speed up the calculation, it
took around 20 minutes for $30$ different $U$, $U=1,...,30$
 with $M_{^{\Omega }}=128$.  CQA results are
in good agreement with the exact results for small $U$($U<4$) and large $U$($U>15$),
 while GW fails for $U>10$.

\subsection{Comparison with exact diagonalization (2D)}

\begin{figure}[tbph]
\centering
\includegraphics{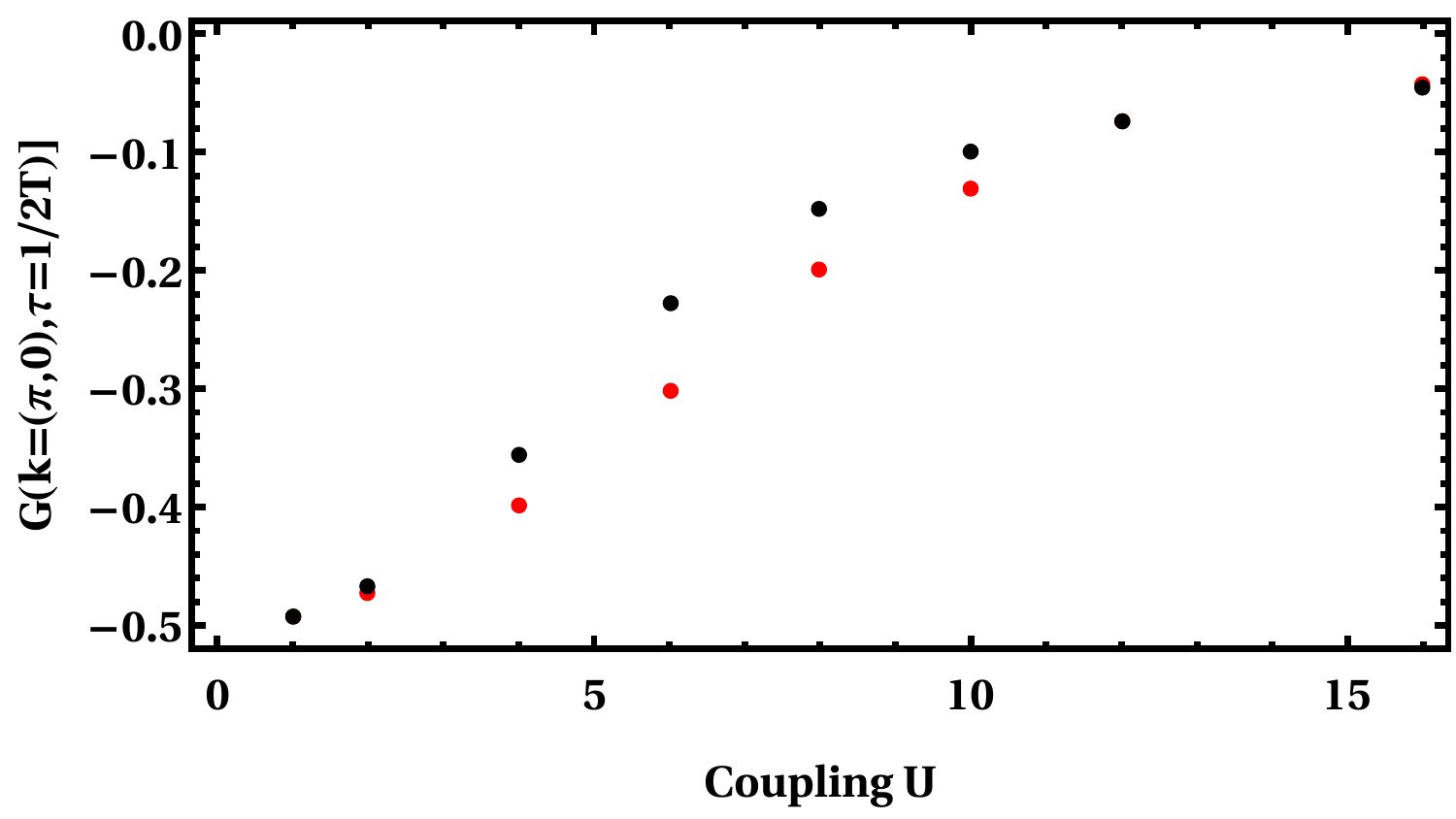}
\caption{Green's function $G(\mathbf{k},\protect \tau )$ of a half filled 2D
Hubbard model at $\mathbf{k}=(\protect \pi ,0)$ and $\protect \tau=1/(2T)$.
The red dots are MC result and the black ones are CQA result.}
\label{fig:2DgvsU}
\end{figure}

We calculated Green's function of 2D Hubbard model, with $N=4$ and $T=1$ at
half filling, for which reliable Monte Carlo simulations exist. The CQA and
Monte Carlo results are shown in Fig.\ref{fig:2DgvsU}. For $U \le 2$ and
$U \ge 12$ CQA results are in perfect agreement with QMC results.

\section{Estimate of complexity of the CQA computation}

\label{sec:complexity}

\subsection{Complexity of general calculation}

\begin{figure}[tbp]
\centering
\includegraphics[width = \linewidth]{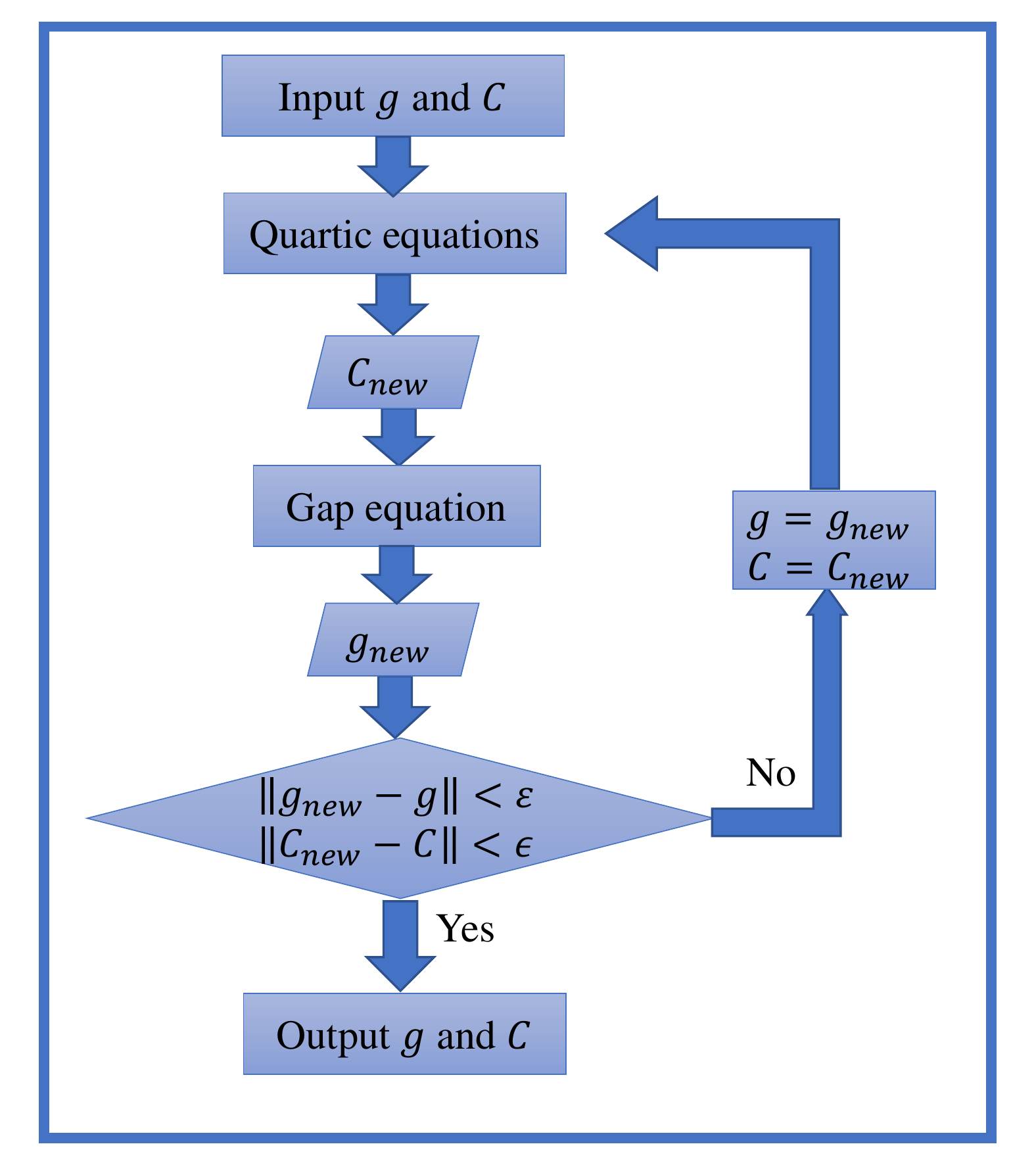}
\caption{Flow chart of CQA calculation}
\label{fig:flowchart}
\end{figure}

In this section we estimate the complexity of a realistic strongly coupled
condensed matter physics model computation using the Hubbard model as a
representative example. Generalization to other models is briefly discussed
in the concluding Section. It is convenient to consider the chain functions
as a (spin/charge/band) vector $C_{\alpha \beta }^{f}$, since the most
computationally intensive part of the computation is the chain equations
iteration. For example in the single band Hubbard model the diffuson and the
cooperon chains can be combined as:$C_{\alpha \beta }^{l}=\left \{ d_{\alpha
\beta }^{1},d_{\alpha \beta }^{2},c_{\alpha \beta },r_{\alpha \beta
}\right
\} $.The computation process is shown in Fig.\ref{fig:flowchart},
where $\varepsilon = 10^{-7}$ and $\epsilon = 10^{-6}$, which leads to the
deviation of the Green function solution less than $0.01 \%$.

Although in the simplest one band local Hubbard model the ``channel" index $%
l $ takes $L=4$ values, the number $L$ becomes larger in multi - band Hubbard
model or  $t-J$ model\cite%
{dagotto1994correlated,scalapino2012common, Fradkin}. The frequency,
quasi-momentum index, $\alpha =\left \{ \omega ,k_{1},...,k_{D}\right \} $,
takes
\begin{equation}
n=2 M_{^{\Omega}} N^{D}  \label{ndef}
\end{equation}
values.\ In addition to chains one also iterates much smaller set of
correlators $g_{\alpha }$. Generally the iteration of $C_{\alpha \beta }^{l}$
involves convolutions of $g$ and $C$ described in detail in Section \ref%
{sec:implement}.

The computational cost of one iteration is estimated as follows. Computation
for fixed $\beta $ and $L$ consists of roughly $L$ times convolutions
implemented for example in MKL\cite{MKL}\ (that as is done using fast
Fourier transform\ and require $n\log \left[ n\right] $ operations each\cite%
{NR}). Different $\beta $ are computed in parallel. For a cluster with
number of cores $N_{cores}<n$, the calculation time therefore can be
estimated as
\begin{equation}
C=\frac{Ln^{2}}{N_{cores}}\log \left[ n\right].  \label{complexity}
\end{equation}

\subsection{Our calculation}

In our $D=2$ sample computation for the simplest model described above one
has: $L=4,$ $M_{^{\Omega}}=128,$ $N=4$ so that $n=256\times 4^{2}=4096$.
The frequency cutoff $M_{^{\Omega}}$ is sufficient to simulate a relatively large temperature $%
T=t $. For the hopping parameter $t=0.4eV$ this amount to $4600K$. More
relevant temperature range of $100-1000K$ will require larger $M_{^{\Omega}}$. To
determine frequency cutoff at any temperature and coupling the
convergence of $M_{^{\Omega}}\rightarrow \infty $ was studied. The energy cutoff
($M_{^{\Omega}}T$) should be larger than relevant energy scales: band width
(hopping energy) and the on site Coulomb repulsion energy $U$ (coupling).
The frequency cutoff is therefore estimated as,%
\begin{equation}
    M_{^{\Omega}}=8 T^{-1}\max \left[ 2t,U/2\right] \text{.}
\label{frequencyestimate}
\end{equation}

The present work uses $N_{cores} = 32$ cores on the High-performance
Computing Platform of Peking University.
The equation Eq.(\ref{complexity}) thus amounts to $C=\frac{4\times 4096^{2}%
}{32}\log \left[ 4096\right] =2\times 10^{7}$ operations per iteration. At
not very large coupling used for the present exploratory calculation \
(largest coupling to temperature ratio $U/T=30$) the number of iterations
required for convergence $N_{it}$ scales with $U$ as
\begin{equation}
N_{it}=10 U/T.  \label{iternumber}
\end{equation}%
%
In our calculation each iteration took $16$ seconds, 
and 30 values of coupling $U=1,2,\cdots,30$ for a single value of
temperature and frequency cutoff were calculated, it over all took 21 hours.

For a realistic applications(still in 2D but for a several band Hubbard type
model and quasi - local interactions and higher values of coupling), more
powerful computational resources are required.

\subsection{An estimate for a realistic 2D system.}

For a more realistic applications one uses room temperature $T=300K$
(sometimes below for example when high $T_{c}$ superconductivity is
considered the relevant range is below $100K$). Therefore it is feasible to
perform calculation for strong coupling $U/T=80$ required for $T=0.1t$.

For a popular 3 band Hubbard model with several couplings considered to
realistically describe perovskite 2D materials\cite{dagotto1994correlated,
scalapino2012common} number of channels is about $L=10$. The values of the
hopping parameter is about $t=0.25eV$ and the typical coupling strength $U=8t
$. According to estimate Eq.(\ref{frequencyestimate})\ $M_{^{\Omega}}
=32t/T=320$. Monte
Carlo simulations at half filling \ demonstrate that the continuum limit in
this case is achieved for $N_{s}=12$. The number of cores available in
Normal\ Taiwan University is $N_{cores}=1080$.
The number of ``degrees of freedom" is consequently $n=1\times 10^{5}$.
According to the estimate Eq.(\ref{complexity}), the one iteration time is
determined by $C=\frac{10\times \left( 1 \times 10^{5}\right) ^{2}}{1080}%
\log \left[ 1. \times 10^{5}\right] =1.5\times 10^{9}$ operations. This
amounts to $33$ minutes.

The maximal number of iterations required estimates $N_{it}=10\times 80=800$
. The computation of the two - body correlator will take two weeks.

\section{Discussion and conclusions}

\label{sec:discussion}

To summarize, we have developed a non - perturbative manifestly charge
conserving method, covariant quartic approximation, determining the
excitation properties of crystalline solids.  It was shown that truncations
of the set of Dyson - Schwinger equations for correlators of the downfolded
model of materials lead to a converging series of approximants. The
covariance ensures that all the Ward identities expressing the charge
conservation are obeyed. A large number of solvable bosonic and fermionic
field theoretical models demonstrate that the approximant in this series, is
sufficiently precise. We focus here on the electron correlators describing
single electron (hole) excitations observed directly by for example the
photo-emission experiments.

The scheme was implemented on supercomputer and tested on the one band
Hubbard model in both 1D (where exact diagonalization results were derived)
and 2D at half filling (where determinantal quantum Monte Carlo results exist%
\cite%
{blankenbecler1981monte,hirsch1985two,santos2003introduction,ma2013quantum,
ma2018localization}). Estimates of the complexity for more realistic lattice
models like local multi - band Hubbard were made.

The method is applicable in cases where various Monte Carlo based methods
fail due to the sign problem\cite{MC}. In most cases of interest the sign
problem therefore does not allow simulation. When comparing the two approaches
in the absence of the sign problem (like in the half filled Hubbard model),
note that the value of the frequency cutoff used in quartic approximation is
much larger than the corresponding value used in MC simulations to achieve
same precision. The MC estimate of the cutoff is $M=T^{-1}\sqrt{8tU}$, where
$U$ is strength of interactions, $t$ is hopping parameter (band width for
narrow bands) \ and $T$ is temperature. This should be compared with our
estimate in Eq.(\ref{frequencyestimate}).

For realistic applications to strongly correlated electronic systems\cite%
{Fradkin}, more complicated models on the ``mesoscopic scale" should be
considered. Although the basic band structure of crystalline solids can be
theoretically investigated by the density functional methods, the condensed
matter characteristics dependent on the detailed structure of the electronic
matter near the Fermi level requires more precise treatment of the relevant
degrees of freedom conveniently represented as an ``effective" lattice model
on the scale of nanometer often taking the form of the local multi - band
Hubbard model or a quasi - local $t-J$ model.
Like some other methods for example versions of GW \cite{Gunnarson},
various Monte Carlo based methods like DMFT\cite%
{Kotliar},  FLEX\cite{FLEX1,FLEX2} and parquette\cite%
{rubtsov2012dual,astretsov2019dual}, CQA therefore can also be applied
to the effective lattice models with pairwise interactions.

\begin{acknowledgments}

Authors are very grateful to T. X. Ma for providing the Monte Carlo data and
J. Wang, H. C. Kao, for numerous discussions and help in computations. B.R. was supported
by MOST of Taiwan, Grants No. 107-2112-M-003-023-MY3. D.P.L. was supported by National
Natural Science Foundation of China, Grants No. 11674007 and No. 91736208.
B.R. and D.P.L. are grateful to School of Physics of Peking University and The Center for Theoretical Sciences
of Taiwan for hospitality, respectively.

\end{acknowledgments}
\bibliography{QAHubbard}

\end{document}